\documentclass{aa}

\pdfoutput=1 
\usepackage[varg]{txfonts}
\usepackage{natbib}
\def\gta{\;\lower 0.5ex\hbox{$\buildrel > \over \sim$}}
\def\lta{\;\lower 0.5ex\hbox{$\buildrel < \over \sim$}}
\usepackage[section]{placeins} 
\titlerunning{dSph: Gravitational tide effects}
\authorrunning{Nichols~et~al.}

\begin{document}

\title{The effect of gravitational tides on dwarf spheroidal galaxies}
\author{Matthew~Nichols \inst{\ref{inst1}\thanks{\email{matthew.nichols@epfl.ch}}}
\and
Yves~Revaz \inst{\ref{inst1}}
\and
Pascale~Jablonka \inst{\ref{inst1},\ref{inst2}}}

\institute{Laboratoire d'Astrophysique, \'Ecole Polytechnique F\'ed\'erale de Lausanne (EPFL), 1290, Sauverny, Switzerland \label{inst1}
    \and
    GEPI, Observatoire de Paris, CNRS UMR 8111, Universit\'e Paris Diderot, 92125, Meudon Cedex, France \label{inst2}}
\date{}
\abstract{The effect of the local environment on the evolution of dwarf spheroidal galaxies is poorly understood.
We have undertaken a suite of simulations to investigate the tidal impact of the Milky Way on the chemodynamical evolution of dwarf spheroidals that resemble present day classical dwarfs using the SPH code GEAR.
After simulating the models through a large parameter space of potential orbits the resulting properties are compared with observations from both a dynamical point of view, but also from the, often neglected, chemical point of view.
In general, we find that tidal effects quench the star formation even inside gas-endowed dwarfs.
{Such quenching, { may} produce the radial distribution of dwarf spheroidals from the orbits seen within large cosmological simulations.}
We also find that the metallicity gradient within a dwarf is gradually erased through tidal interactions as stellar orbits move to higher radii.
The model dwarfs also shift to higher $\langle$[Fe/H]$\rangle$/L ratios, but only when losing $\gta$$20\%$ of stellar mass.
}

\maketitle
\section{Introduction}
Simulations of the prevailing cosmological {model} -- dark energy and cold dark matter, $\Lambda$CDM -- predict that a copious amount of dark matter subhalos {will} survive to the present day.
Within the Local Group these subhalos will be of a mass comparable to the halos that surround dwarf galaxies.

Although there is an underabundance of observed dwarf galaxies compared to the amount of subhalos predicted by $\Lambda$CDM \citep[e.g.][]{Klypin1999,Moore1999,Sawala2010}, numerous discoveries in the last decade \citep{Willman2005,Willman2005a,Zucker2006,Belokurov2007,Irwin2007,Belokurov2008,Belokurov2009,Belokurov2010,Willman2011} and a greater understanding of baryonic processes has somewhat alleviated this \citep{Bullock2010,Sawala2011}.
All of these recently discovered dwarfs, and a majority of the previously known Local Group satellites are low luminosity, gas-deficient dwarf spheroidals (dSphs).
Their low mass; proximity to the Galaxy (allowing individual stars, albeit only the brightest ones, to be resolved); and that they represent the final evolutionary state of many (though not all) dwarf galaxies \citep{Grebel2003}; has made dSphs observational targets for studying a wide variety of processes related to galaxy formation.

Dynamical studies of dSphs have been undertaken by calculating the dynamical history of the Local Group dSphs and from numerical calculations determining the allowed phase-space of dSphs.
Using the observed proper motions of dSphs \citet{Lux2010} calculated the allowed perigalacticon and apogalacticon of seven Milky Way (MW) dSphs (and the Magellanic Clouds) by integrating the orbits backwards in time through a static potential.
By calibrating the results to results from the Via Lactea II simulation they conclude that the MW's dSphs are likely to have formed before reionization.
Importantly, the recorded dSphs with the lowest perigalacticon correspond to those with tidally induced inner bars \citep{Lokas2012}.
Similar tracing of orbits backwards in time has been undertaken by others for individual galaxies \citep[e.g.][]{Pasetto2011} finding similar results.
In all cases the uncertainty in the proper motion and the mass accretion history of the MW limits the accuracy of the resolved orbit.

In contrast to determining the orbital history of the dwarfs by integrating the observed motion backwards in time, \citet{Rocha2012} compared the present day {kinematical} properties of dSphs with the Via Lactea II simulation to determine a likely accretion time for these dwarfs.
This accretion time (aka the infall time, the time at which a satellite galaxy first crosses the virial radius of its host) is compared to the star formation history of the dwarf galaxies.
Here, the ultrafaint dSphs are found to have likely ceased star formation before infall.
For some classic dSphs there is an overlap between possible accretion times and dramatic increases in the star formation before being quenched within approximately a Gyr of infall.
That such overlaps exist for a few of the classical dwarfs (although not all, Carina for instance shows a burst of star formation post any likely infall time) suggests the environment strongly influences the star formation rate of dSphs, potentially being responsible for the quenching of dwarf galaxies \citep{Rocha2012}.

Complementing the orbital history calculations of dSphs, some authors have examined the effect of interactions between the MW and the orbiting dSph.
Dynamically the two major components of these interactions are tidal forces arising from the dSphs interaction with the gravitational field of the MW, and gas stripping due to the interaction of the dSph with the predicted hot gas halo of the MW \citep{Blitz2000,Grcevich2009,Miller2013,Gatto2013}.

Examining the tidal forces acting upon simulated dwarfs, \citet{Penarrubia2008} found that many dwarf galaxies will be able to retain a stellar density profile of similar shape to their original profile, even when retaining only a small fraction ($\sim$$1\%$) of their stars.
Due to the "cuspy" nature of dark matter protecting it from stripping, such dwarfs have evolved elevated dynamical to luminous matter ratios, similar to that of dSphs.
Consequently, the maximum circular velocity of the dwarf was found to drop in line with the decreasing stellar mass.
This dependence is not seen inside the local dSphs and hence \citet{Penarrubia2008} found it unlikely that dSphs are tidally evolved remnants from more massive dwarf galaxies.

{\citet{Kazantzidis2013} explored the effect of the dark matter potential on this stripping and found that dwarfs with a less "cuspy" profile are more likely to be transformed into systems reminiscent of dSphs.
In particular for these core-like dwarfs, such transformation can take place within one pericentric passage and dwarfs on near circular orbits and at large perigalacticons are able to be transformed.
When considering which dwarfs are destroyed, they find that dSphs are more likely to be found on circular orbits or at higher perigalacticons.
With eccentric, low perigalacticon dwarfs ending up as tidal streams.}

\citet{Mayer2006} examined both tidal and ram pressure stripping through high resolution {\em N}-body+SPH simulations of gas rich dwarfs (sans stellar feedback or chemical evolution) within a MW like halo.
Tidal stirring of the dwarfs produces objects with kinematics resembling dSph while ram pressure stripping removes the gas at low perigalacticons.
Such processes were sufficient to strip low mass dwarfs with low perigalacticons and with ultraviolet heating from the extragalactic UV field (or what may be expected from stellar feedback), stripping of more massive dwarfs at similar perigalacticons.
That ram pressure stripping is sufficient to remove this gas at low perigalacticons agrees with analytic predictions for dwarf galaxies \citep{Mori2000}.
It is important to note however, that SPH simulations, such as those undertaken by \citet{Mayer2006} suffered from numerical problems \citep[see][]{Agertz2007} due to the artificial viscosity present in standard SPH formulations which enhanced the survivability of gaseous clouds against stripping and predictions concerning ram pressure stripping may suffer from some inaccuracy as a consequence of this.

Although tidal and ram pressure stripping are sufficiently powerful to strip galaxies close in (perigalacticons$\lta50$~kpc) to the host galaxies, the gas deficiency of dwarfs continues well past $200$~kpc \citep{Grcevich2009}.
At these large distances, stellar feedback may provide sufficient heating to allow gas to be readily stripped even by the sparse {hot gas} halo \citep{Nichols2011,Nichols2013}.
Understanding how the star formation history of dwarf galaxies is therefore immensely important in understanding how they evolve over cosmic time.

Parallel to studies of dynamical evolution, much work has been conducted trying to understand the chemical evolution of dwarf galaxies.
With evolutionary histories that often contains periods of heightened star formation after the initial burst \citep[e.g.][]{Tolstoy2009,Lee2009} and large metal loss which has been proposed analytically to be expelled through supernova driven winds \citep{Kirby2011b}.
These heightened periods, seen in both isolated and tidally interacting dwarfs \citep[{although such dwarfs are potentially more massive than the dSphs today as seen by their ability to retain gas}]{Brosch2004} may provide an environment where metals may quickly pollute the ISM before the next period of star formation commences, however, how quickly the metals are able to be redistributed and the quantity {lost during the burst} remain open questions.

We have previously examined how the star formation history imprints on the chemical abundances of dwarf galaxies through chemodynamical processes in isolated dwarfs using the TreeSPH code GEAR \citep{Revaz2009,Revaz2012}.
The code fully solves the gravitational and hydrodynamics in addition, it accounts for star formation processes, chemical enrichments, cooling and supernova feedback.
\citet{Revaz2009,Revaz2012} obtained generic models that described the great variety of dSphs observed in the local universe.
The mass ranges and star formation histories of these generic models are compatible the with observed dSphs like Carina, Sextans, Sculptor and Fornax.
{However, star formation continued on past the initial burst, preventing the development of dwarfs with Sextans or Sculptor like star formation histories which display a single, terminated, period of star formation \citep{LeeMG2009,deBoer2012}.
In order to fully match the present day chemical evolution, this star formation had to be artificially terminated for Sextans and Sculptor.
As these dwarfs relied only on internal processes, the requirement of artificial termination is in line with previous work, with \citet{Sawala2010} demonstrating that internal feedback is not sufficient to reproduce the full range of observed galaxies.
In particular to reproduce dwarfs that only have a short burst of star formation initially before it infall onto a larger galaxy, additional UV heating (such as from the extragalactic UV field) {may be} necessary \citep[see also e.g.][]{Ricotti2005,Valcke2008}.}

{Even with the artificial termination in star formation history reproducing the observed metallicity, mass and luminosity}, the model galaxies retained large quantities of gas ($\gta10^{7}~M_{\sun}$) to the present day, in conflict with the gas deficiency observed {in satellite galaxies within the Local Group} \citep{Grcevich2009}.
However, the closest dwarf galaxies (and subsequently the ones we are best able to study) are -- by virtue of their proximity -- not isolated.
How the interaction between a low mass galaxy and an MW sized galaxy affects the chemical abundances inside the dwarf galaxy is currently unknown, with no previous studies exploring this.
Importantly, is it possible to have the gas removed from dwarfs just through this interaction without greatly modifying the chemical or star formation properties of a dwarf, or are calculations of dwarfs in isolation unsuitable for the nearest dwarfs{?}
Here we examine whether tidal interactions between model dSph and the MW greatly alter the chemodynamical properties of the dSph and what this may reveal about the progenitor properties of some of the classical dwarf spheroidals.
In \S\ref{sec:IC} we describe {the GEAR code}, model setup and initial conditions of the model dwarfs, the impact of tidal forces are discussed in \S\ref{sec:tide}, with concluding remarks in \S\ref{sec:conclusion}.

\section{Model}\label{sec:IC}

Modelling the effect of tides requires modelling both the orbit of the dSph as it moves through the hot halo of the MW in addition to realistic models of the gas and dark matter profiles of both objects.

\subsection{GEAR}

{In order to accurately calculate the chemical evolution of dSphs as they undergo fluctuating tidal forces throughout their orbit we use the code GEAR \citep{Revaz2012}.
GEAR is a TreeSPH code supplementing the public version of Gadget-2 \citep{Springel2005a} with a complex treatment of the baryonic physics.
Its excellent behaviour with respect to integral conservation as well as its convergence with increasing resolution has been previously demonstrated.

In GEAR, we take into account the radiative cooling of the gas. Above $10^4$~K we use the metallicity-dependent prescription of \citet{Sutherland1993}, while below, we take into account the cooling due to H$_2$ and HD molecules as well as the cooling due to fine structure line transition of the most important metals, oxygen,carbon, silicon, and iron \citep{Maio2007}.
We presently ignore the UV background radiation heating which can be neglected at the resolution of these simulations where overdensities that would efficiently radiate energy \citep{Omont2007} are not sufficiently resolved.
Higher resolution rests showed negligible differences to the final state of the dwarf.

Stars forms from the dense ($\rho_{\rm{sfr}}>0.1~\rm{atom~cm}^-3$) cold ($T<3\times 10^4~\rm{K}$) and contracting gas, following the stochastic approach
initially proposed by \citet{Katz1992,Katz1996}.
Each gas particle may form up to 4 stellar particles \citep{Springel2003}, each representing a single stellar population (sharing the same age and metallicity).
We use the IMF modelisation from \citet{Kroupa2001}.

For every timestep following the formation of a stellar particle, we precisely compute the number of Type II (SNII) and Type Ia (SNIa) supernovae exploding
during this timestep. Energy and synthesised elements are injected among the 50 nearest neighbors, following the SPH scheme.
For each exploding supernova { (of both types)}, we assume that only $3\times10^{49}$~erg of thermal energy impacts the ISM, while the remaining is assumed to leave the system, due to radiative cooling occurring in the shock front which is not resolved here.
In order to avoid an instantaneous cooling of the hot gas impacted by this energy, we use an adiabatic time $t_{\rm{ad}}=5~\rm{Myr}$ during which the cooling of those particles is artificially switched off \citep{Stinson2006}.
The abundances of Fe and Mg are primarily followed and their yields are taken from \citet{Tsujimoto1995} (SNII) and \citet{Iwamoto1999} (SNIa).

The parameters used for the different dSph simulations described further are detailed in Table \ref{tab:dSphparam}.
}

\subsection{Dwarf Spheroidal Orbits}
The orbits of Carina, Sextans, Sculptor and Fornax have been studied numerous times over the last decade \citep[e.g.][]{Piatek2002,Piatek2003,Piatek2006,Piatek2007,Lux2010,Rocha2012}.
Carina, Sextans and Sculptor possess similar orbits with perigalacticon distances of $\sim40$--$80$~kpc and apogalacticons of $\sim100$--$250$~kpc.
This similarity between the orbits allows a convenient suite of simulations.
Fornax, due to stricter constraints suggesting a more circular orbit, is treated separately with a perigalacticon of $\sim$$120$~kpc and apogalacticons of $170$--$250$~kpc.

Each dSph will experience the model MWs gravitational potential, from the time it is captured until the present day.
This time is constrained by the present day position of the dSph and the perigalacticon/apogalacticon combination.
The orbital period ($T^{a,p}_{\rm orb}$, where $a$ is the apogalacticon in kpc and $p$ the perigalacticon in kpc) is sufficiently large that the dSphs are limited to a maximum of three perigalacticon passes within the age of the Universe.

{We supplemented Gadget-2 with the possibility of adding an external potential in order to model these orbits.
The presence of such an external potential results in two major simplifications to the orbits of the dSphs.
Firstly, the external potential is static in time, this will have the effect of altering the tidal forces experienced early on in the dSphs history and may result in an exaggeration of results for those which have experienced multiple orbits.
Secondly, dynamical friction is ignored, but in the mass regime of dwarf galaxies, this will only have a minor impact due to the large mass contrast meaning dynamical friction will have only a minor impact upon the orbits.}

As dSphs will likely be near their apogalacticon when accreted, the simulation is begun there and continued until the dwarf reaches its current position given in Table \ref{tab:dSphpos}.
A suite of simulations is performed for which dSph experience $1$, $2$, or $3$ perigalacticon passages ($n_{\rm peri}$).
The time in which a dSph experiences the influence of the host galaxy is then at least $T_{\rm MW} = \left(n_{\rm peri}-\frac{1}{2}\right)T^{a,p}_{\rm orb} + T_{p2c}$, where $T_{p2c}$ is the orbital time between the perigalacticon and the current position.

\begin{table}
  \caption{{Observed current positions of the four dSph addressed}}
  \label{tab:dSphpos}
  \centering
  \begin{tabular}{lcccl}
    \hline\hline
{dSph}&{$X$}&{$Y$}&{$Z$}&{Reference}\\ &{[kpc]}&{[kpc]}&{[kpc]}&\\
\hline

  Carina&$2.4$&$91$&$-38$&[1]\\
  Sextans&&$r_0=85$&&[2]\\
  Sculptor&$5.4$&$9.9$&$-86$&[1]\\
  Fornax&$39$&$48$&$-126$&[1]\\
\hline
\end{tabular}
\tablefoot{$(X,Y,Z)$ coordinates correspond to a right-handed coordinate system with an origin located at the MWs centre. The disk of the MW coincides with the $X$-$Y$ plane, with the present location of the Sun on the positive $X$ axis. {Sextans is listed only as a current-day radial distance $r_0$. As the MW is treated as a spherical object in this paper, this has no effect on any results.} { References. [1] \citep{Piatek2003}; [2] \citet{Irwin1990}.}}
\end{table}

In order to simulate a dSph over the lifetime of the universe ($T_U$ set to $14$~Gyr for simplicity), the dSph should be evolved in isolation for a time $T_{\rm iso}=T_{U} - T_{\rm MW}$.

The simulation parameters are detailed in the appendix (Tables \ref{tabA:paramCar}--\ref{tabA:paramFnx}).
The seemingly arbitrary perigalacticons arise from a programming error in the initial orbit calculations to set the starting time.
Of course, this does not affect the results in any form, with the suite still covering the desired range of perigalacticons, apogalacticons, and orbital periods allowing a full study of the environmental impact on dSph evolution.

\subsection{Profiles}

\subsubsection{Dwarf Spheroidals}

The initial profile of the dSph consists of dark matter and gas particles distributed following a pseudo-isothermal profile
\begin{equation}
  \rho(r) = \frac{\rho_{\rm c}}{1+r^2/r^2_{\rm c}},
\end{equation}
where $\rho_{\rm c}$ is the core density and $r_{\rm c}$ is a core scale radius.
The initial gas profile was set to be equal to the dark matter profile scaled to give the universal gas/total mass fraction [$\rho_{{\rm gas}}(r)=0.15\rho_{{\rm tot}}(r)$].
{The choice of a pseudo-isothermal profile is supported by the observations of low brightness dwarf galaxies \citep[e.g.][]{deBlok2008,Spano2008,Oh2011,Walker2011}, although such a profile is likely to enhance tidal effects \citep{Kazantzidis2013}.}

{The gas temperature was calculated by assuming it to be in hydrostatic equilibrium in the potential well of the dark halo.
The initial velocities of the gas particles are set to zero while the velocities of the dark halo particles are set according to the jeans equations in spherical coordinates.
{ We previously ran several simulations on similar mass halos with UV background radiation. However we see only negligible effects due to this radiation. Therefore, we do not consider a UV background in this paper.}}

The model values used in the simulations are shown in  Table \ref{tab:dSphparam}, and were chosen from the best models obtained by \citet{Revaz2012}.
In addition to the standard resolution models, we also run isolation models and one interaction at a higher resolution (models with suffix HR) to investigate both the effects of resolution on the results.

\begin{table*}
  \caption{Simulation parameters for the four generic dSph models}
  \label{tab:dSphparam}
  \centering
  \begin{tabular}{lcccccccc}
    \hline\hline
{dSph}&{$r_{\rm max}$}&{$c_\star$}&{$\rho_{\rm c,gas}$}&{$M_{\rm tot}$}&{$N_{\rm part}$}&{$M_{\rm DM,part}$}&{$M_{\rm gas,part}$}&{$\epsilon_{\rm g}$}\\ &{[kpc]}&&{m$_{\rm H}~$cm$^{-3}$}&{$10^8~M_{\sun}$}&&{$M_{\sun}$}&{$M_{\sun}$}&{pc}\\
\hline
  car&$3.5$&$0.1$&$0.022$&$1$&$65536$&$3051$&$458$&$50$\\ 
  sex&$8.0$&$0.05$&$0.022$&$3$&$65536$&$9155$&$1373$&$50$\\
  scl&$9.6$&$0.05$&$0.029$&$5$&$65536$&$15258$&$2289$&$50$\\
  fnx&$7.1$&$0.05$&$0.059$&$7$&$65536$&$21362$&$3204$&$50$\\
  carHR&$3.5$&$0.1$&$0.022$&$1$&$524288$&$381$&$57$&$25$\\
  sexHR&$8.0$&$0.05$&$0.022$&$3$&$524288$&$1144$&$172$&$25$\\
  sclHR&$9.6$&$0.05$&$0.029$&$5$&$524288$&$1907$&$286$&$25$\\
  fnxHR&$7.1$&$0.05$&$0.059$&$7$&$524288$&$2670$&$401$&$25$\\
\hline
\end{tabular}
\tablefoot{{ Simulation parameters for generic dSph models}. { The parameters are} chosen to produce the best fitting model to observed dSphs when evolved in isolation, $r_{\rm{max}}$ {is the initial radius}, $c_\star$ is the star formation parameter, $\rho_{\rm{c,gas}}$ is the density of gas at the centre and $M_{\rm{tot}}$ is the total mass, {$N_{\rm part}$ is the { total} number of particles { equally split between gas and dark matter}}, { $M_{\rm DM,part}$ is the dark matter particle mass,$M_{\rm gas,part}$ is the initial gas particle mass, }, and $\epsilon_{\rm g}$ is the gravitational softening length of the dark matter and gas { (the stellar softening length is $1.25{\epsilon_{\rm g}}$)}. See also Table 4 of \citet{Revaz2012}.}
\end{table*}

\subsubsection{Milky Way}

\begin{figure}
  \includegraphics[width=0.5\textwidth]{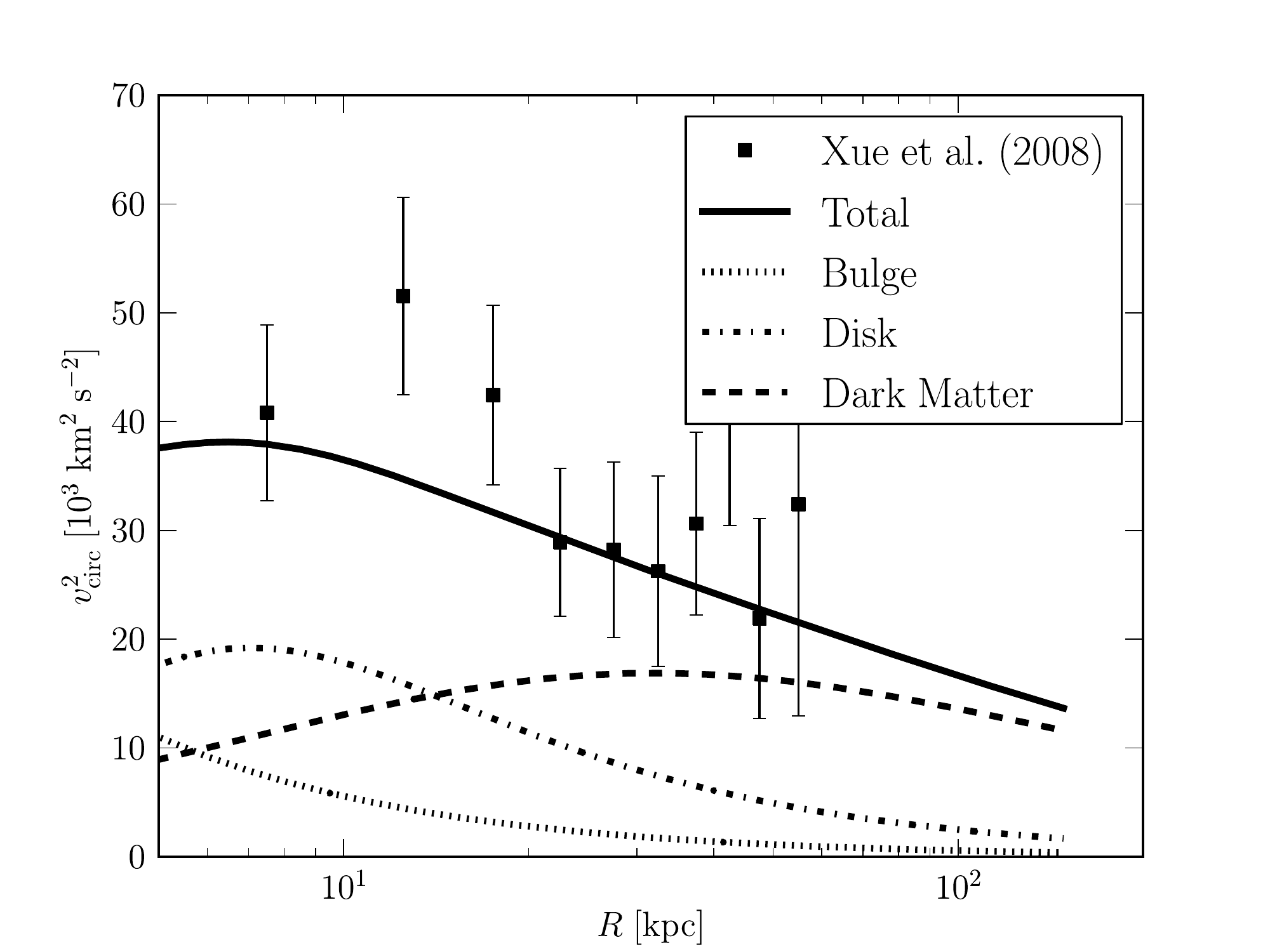}
  \caption{Circular velocity of the chosen Milky Way model. The total rotation curve (solid line) is consistent with the values from \citet{Xue2008}. The contribution of the bulge (dotted line), disk (dash-dot line) and the dark matter halo (dashed line) are also displayed. }
  \label{fig:vcirc}
\end{figure}

The bulge, disk and dark matter halo of the MW are modelled by a fixed potential which reproduces a realistic rotation curve, shown in Fig. \ref{fig:vcirc}.
The potential of the disk is approximated as a spherically symmetric profile to reduce the degrees of freedom present within the simulation.
As we consider only dSphs with perigalacticons exceeding $30$~kpc, this assumption is unlikely to dramatically change any results.
For orbits below this, the assumption of the spherical disk is likely to have a larger impact as a cylindrical exponential profile will have a larger potential gradient as mass is confined inside a smaller volume.
The bulge and the disk are assumed to follow a Plummer profile, with the gravitational potential given by
\begin{equation}
  \phi(R) = -\frac{GM}{\sqrt{R^2+a^2}},
\end{equation}
where $G$ is the universal gravitational constant, $M$ is the scale mass of the bulge/disk \citep[$1.3\times10^{10}~M_{\sun}$ and $5.8\times10^{10}~M_{\sun}$ respectively, see][]{Xue2008}, and $a$ is a scale radius of the bulge/disk \citep[$0.5$~kpc and $5$~kpc respectively, see][]{Xue2008}.

The dark matter halo is represented by a standard NFW profile \citep{Navarro1997}
\begin{equation}
  \phi(R) \propto -\frac{\ln(1+R/a)}{R/a},
\end{equation}
where the constant of proportionality and the scale radius, $a=14.7$~kpc, are chosen to give a concentration of $10$ and a virial mass of $4\times10^{11}$~$M_\sun$.

\section{Impact of tides on dwarfs}\label{sec:tide}

\subsection{Mass and density profiles}\label{ssec:dens}

\begin{figure}
  \includegraphics[width=0.5\textwidth]{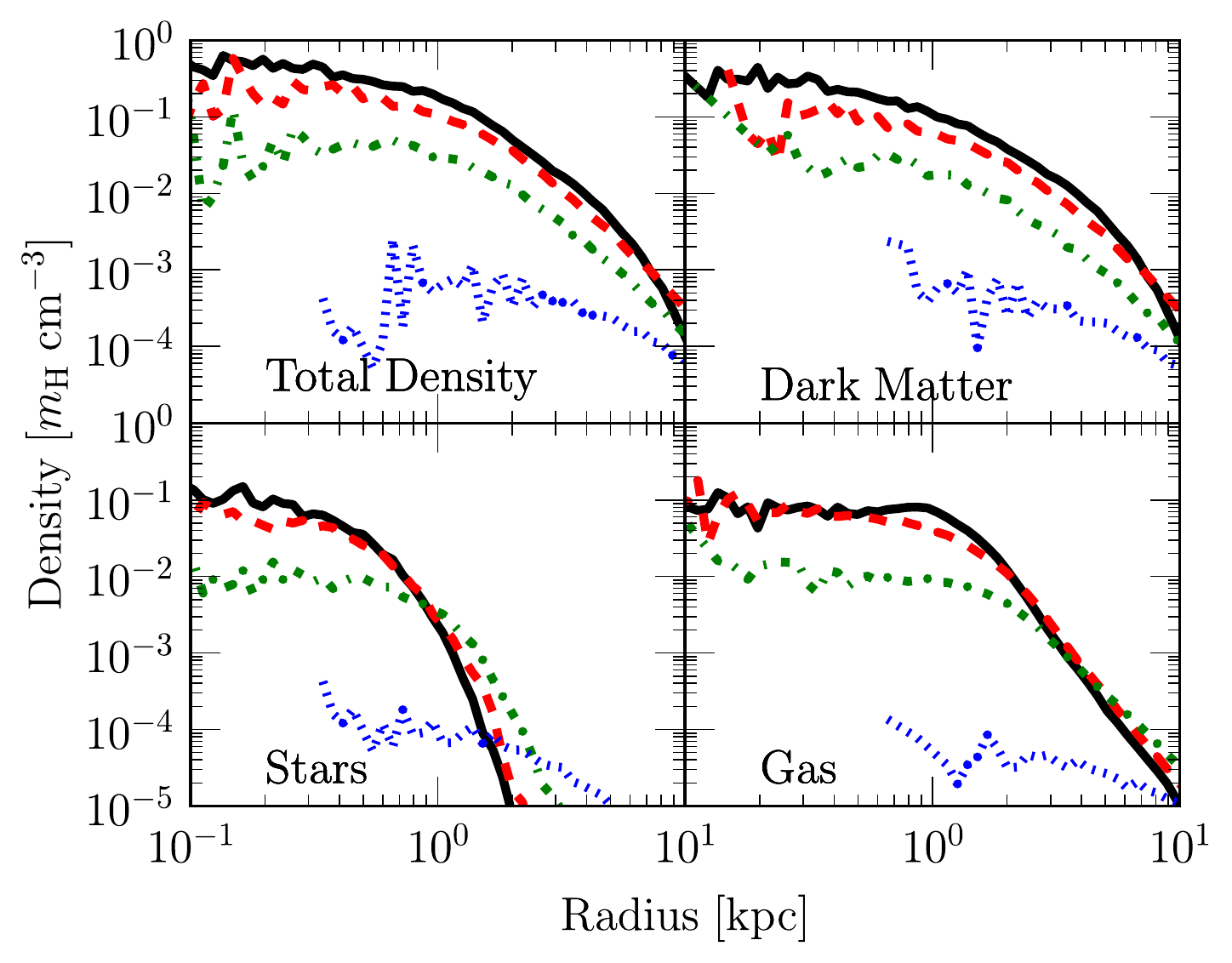}
  \caption{Density profile of a Sextans type dSph with a perigalacticon and apogalacticon of $63$~kpc and $150$~kpc respectively (model sex156). The density is shown at the present time, with the Isolation model shown in a black solid line, models which have undergone one perigalacticon passage marked with a red dashed line, two passages a green dot-dash line and three passages a blue dotted line. Each interaction decreases the central density, allowing tidal forces to more efficiently strip the dSph in future interactions (see also Fig. \ref{fig:sex156mass}). Beyond $10$~kpc each model that has undergone an interaction exceeds the density of the isolation model, indicative of the tidal forces also flattening the profile. \label{fig:sex156dens}}
\end{figure}

\begin{figure}
  \includegraphics[width=0.5\textwidth]{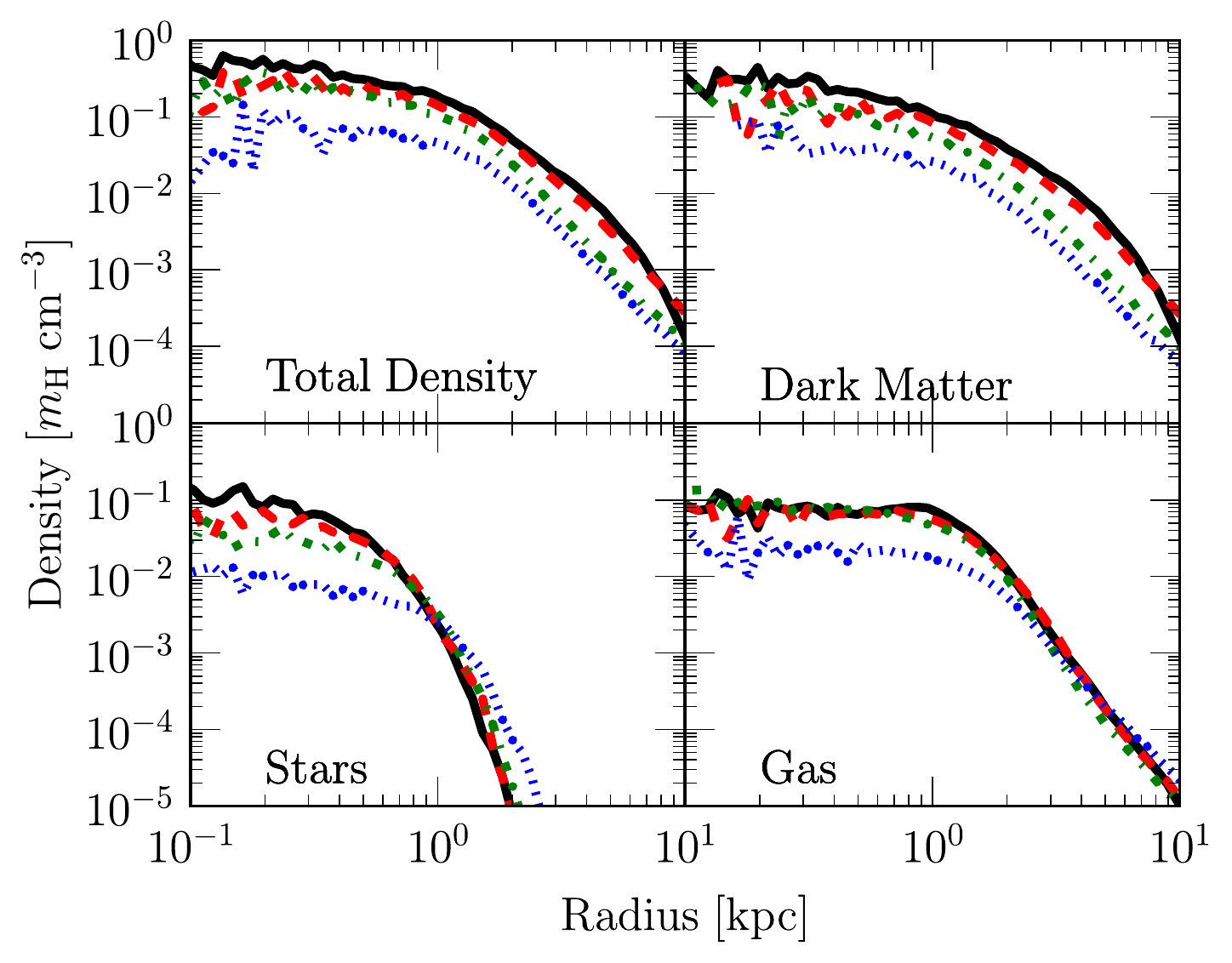}
  \caption{Density profile of a Sextans type dSph with a perigalacticon and apogalacticon of $72$~kpc and $150$~kpc respectively (model sex157). Line colours/styles are the same as Fig. \ref{fig:sex156dens}.
Compared to the model with perigalacticon of $63$~kpc, this model is able to easily survive three perigalacticon passages, producing a final dwarf that is quite similar to Sextans today.  \label{fig:sex157dens}}
\end{figure}

As may be expected the mass loss of dSphs varies dramatically with perigalacticon distance inside the simulations, with dSphs of low perigalacticon experiencing extreme amounts of stripping compared to orbits with large perigalacticons.
In many dSphs, the first passage removes gas and dark matter from the outer boundary of the dwarf, however, such removal is offset by the migration of material from the centre of the dwarf to the outskirts, resulting in a similar, but reduced density profile.
Subsequent passages however, remove a much greater amount of material from the dwarf due to this new, lower, density leaving many almost entirely destroyed and increasing the density scale length in any remaining remnant, particularly at low perigalacticon.
The impact of these tides can be demonstrated by comparing dSphs which are effectively destroyed by tidal stripping and dSphs with slightly larger perigalacticons which survive.
For the three dSphs models whose simulation suites include differing perigalacticons (Carina type, Sextans type and Sculptor type) {--} this threshold, where a dwarf is destroyed but a dwarf at a slightly larger orbit survives {--} occurs around a $60$--$70$~kpc perigalacticon {for the chosen MW and dSph parameters}.
Sextans, which is of a middle mass, is chosen to demonstrate many of these properties, which are applicable to these three sets of models.
We show in Fig. \ref{fig:sex156dens} the impact of multiple passages on the final density profile for model sex156, a Sextans type dwarf of perigalacticon $63$~kpc and apogalacticon of $150$~kpc or various passages and in Fig. \ref{fig:sex157dens} that of the model sex157, a Sextans type dwarf with perigalacticon of $72$~kpc and apogalacticon of $150$~kpc.\footnote{We use the designation dwaABC to correspond to all orbits of passages of dwaABC shown in Tables \ref{tabA:paramCar}--\ref{tabA:paramFnx}. We use the designation dwaABC-n when talking about model dwaABC with n perigalacticon passages. dwa may be Car, for Carina type models; sex, for Sextans type; scl, for Sculptor type; or fnx, for Fornax type. AB represents the first two digits of the apogalacticon, while C represents the first digit of the perigalacticon.}
In both cases, each interaction causes a density drop in the centre, with dark matter and gas migrating from the central regions towards the outskirts.
At these distances one tidal passage is not sufficient to remove much of this material from the dwarf.
However, with two interactions, much more material is able to be removed from the dwarf orbiting at $63$~kpc than the dwarf at $72$~kpc, with dark matter and gas being removed from the centre of the dwarf in the former case and only dark matter a kpc or more away from the centre in the latter case.
Whether two interactions is able to remove material from the centre of the dwarf has a large impact on the survival of models with three perigalacticons.
The dSphs, of all types, with a perigalacticon of $63$~kpc are almost entirely destroyed as the dSph no longer has sufficient density to survive the passage.
The dSphs with three passages of perigalacticon $72$~kpc meanwhile survive mostly intact although with dark matter and gas from the centre stripped from the dwarf, reminiscent of two passage models at the lower perigalacticon.
{In a more massive MW this threshold of survival is likely to occur at larger perigalacticon although few dSphs are likely to have accreted early enough at higher perigalacticons to have undergone three orbits in cosmic time.}

\begin{figure}
  \includegraphics[width=0.5\textwidth]{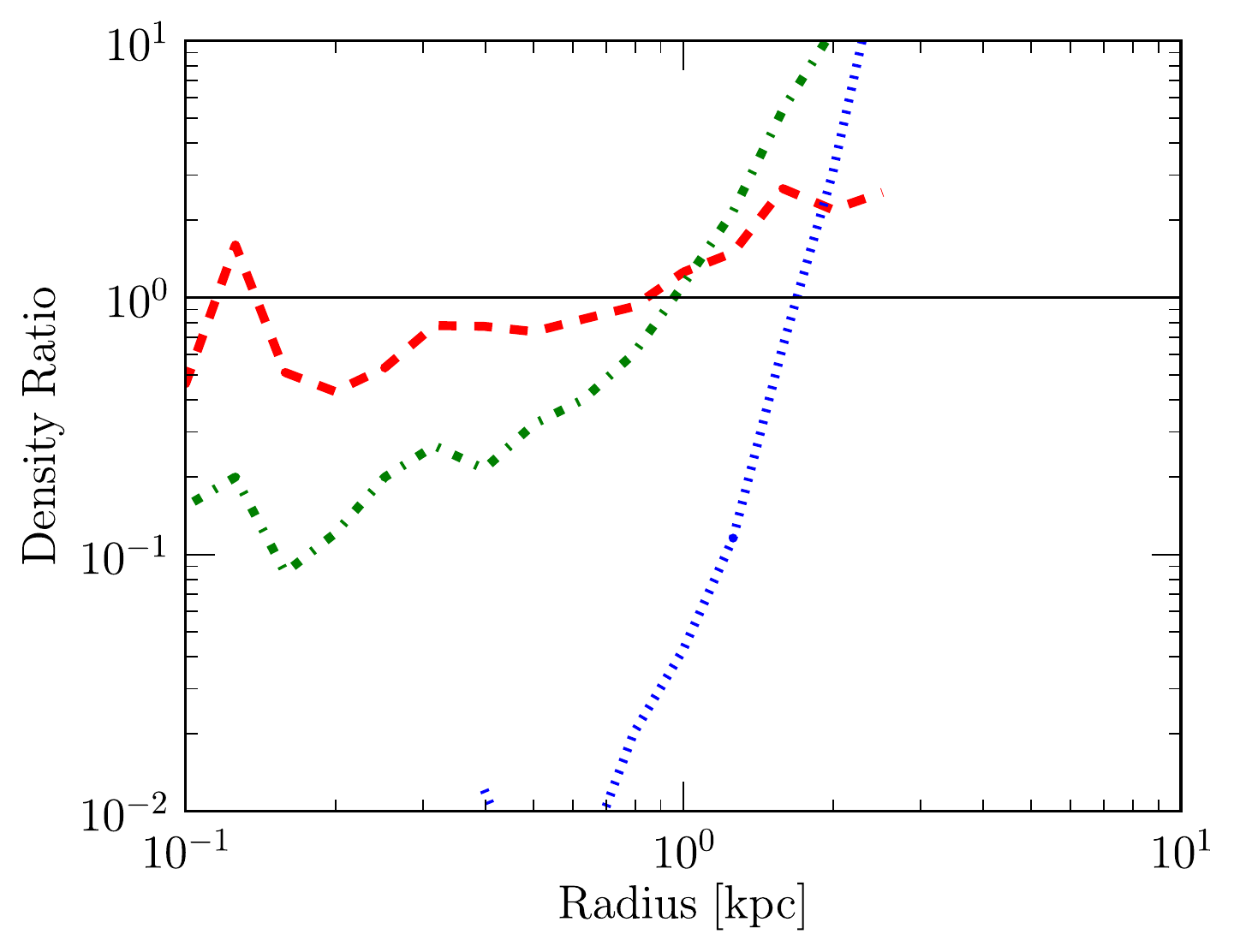}
  \caption{Density ratio of old stars ($>10$~Gyr old) between sex156 models and the Sextans type isolation model. {Line colours and styles are the same as in Fig\ref{fig:sex156dens}.} The age threshold selects only stars that were formed well before star formation was quenched in any simulation (see also Fig. \ref{fig:sex156SFR}). Even with such an old stellar population, the tidal effects are still visible, with all models having the central density lowered (although with some scatter) and increasing in density at higher radius. The magnitude of the change indicates that tidal forces (at $60$~kpc) are sufficient to reduce the central density of a dSph by $80\%$ after two passages and $>99\%$ after three.\label{fig:denscomp}}
\end{figure}

In all cases, the stellar profile is more extended than in real dSphs, with Sersic profiles being a poor fit, particularly in the central region and with scale radii much larger than those observed \citep{Lokas2012}.
That our stellar profile contains a higher proportion of mass in the outside region of the dSphs suggests that in reality the stellar density change may be less prominent.
We take care to note that some of the changed profile in stellar density results however, from the cessation of star formation resulting from perigalacticon passages (see \S \ref{ssec:SFR}).
A quick estimate of the impact of tides (and therefore how much the stellar density changes due to new star formation) can be provided by comparing the density of only old stars in simulations with interactions to that of the isolation model.
We display this density ratio for model sex156 in Fig. \ref{fig:denscomp} with an age cutoff for old stars of $10$~Gyr, selecting stars formed well before star formation has shutoff in any sex156 simulation (about half of all stars formed in the isolation model occur after this point).
Here it is evident that even in this selection, tides are responsible for removing $80\%$ of stars from the central region of the dwarf after two passages and nearly all the stars ($>99\%$) after three passages.
The additional star formation present within models that undergo fewer interactions, see \S\ref{ssec:SFR}, therefore only enhances the difference between models which would still be present if star formation was quenched through another mechanism (e.g. ram pressure stripping).

\begin{figure}
  \includegraphics[width=0.5\textwidth]{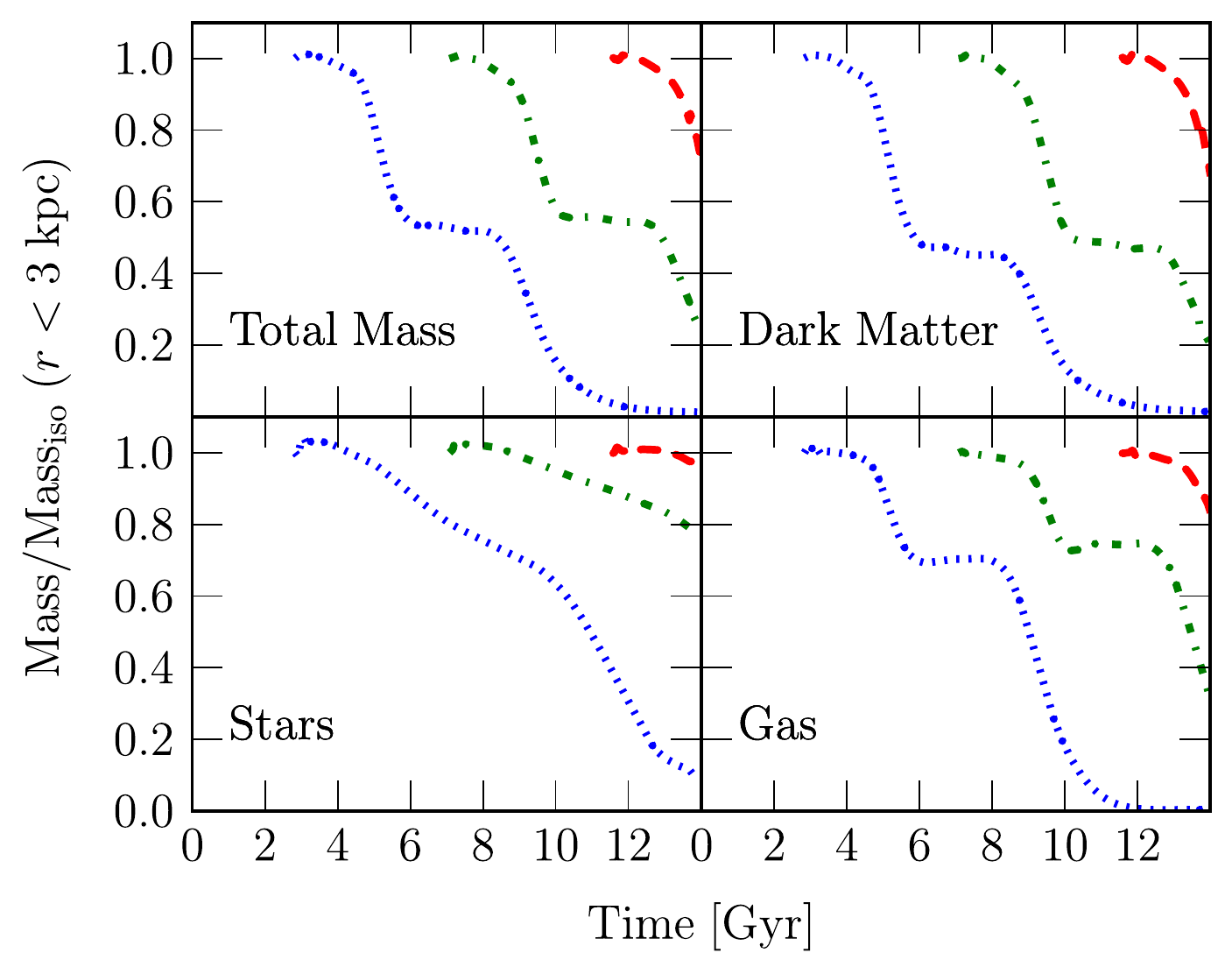} 
  \caption{Ratio of mass contained within $3$~kpc to the isolation model for the dSphs shown in Fig. \ref{fig:sex156dens} (model sex156).
Line colours and styles are as in Fig. \ref{fig:sex156dens}.
The $3$ kpc cut-off displayed is chosen arbitrarily to enclose nearly all of the stellar mass within the isolation model.
Despite the stellar population remaining resilient to tidal stripping, the gas and dark matter is able to be significantly stripped after each interaction resulting in a decrease in the total mass.
The steeper decrease in gas mass -- compared to dark matter -- at approximately $12.5$~Gyr is due to expulsion from supernova feedback before the gas reaccretes at approximately the present day.\label{fig:sex156mass}}
\end{figure}

\begin{figure}
  \includegraphics[width=0.5\textwidth]{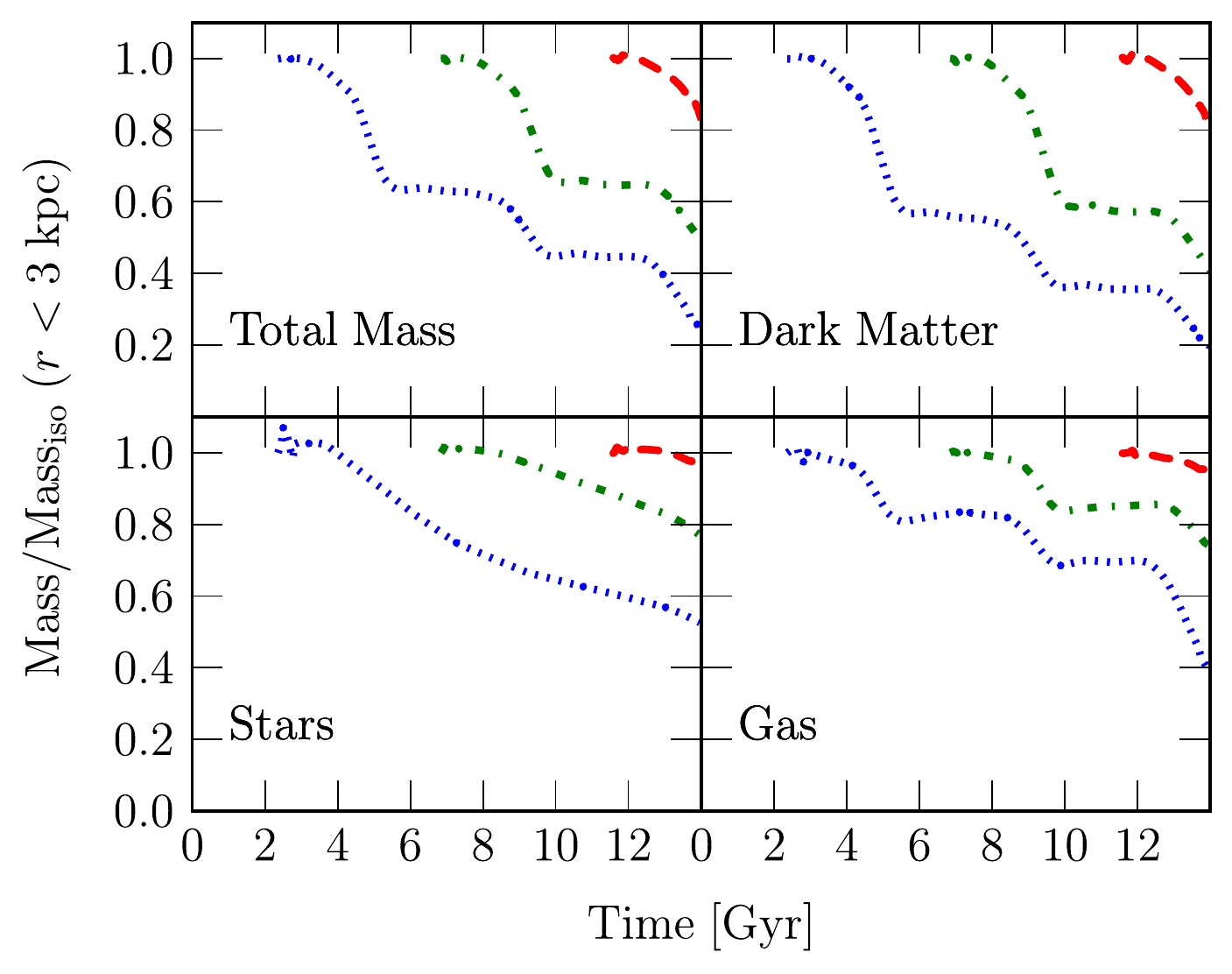} 
  \caption{Ratio of mass contained within $3$~kpc to the isolation model for the dSphs shown in Fig. \ref{fig:sex157dens} (model sex157).
Line colours and styles are as in Fig. \ref{fig:sex156dens}.
Notably, even for three interactions, the gas density remains high enough $(\gta3\times10^{-3}$~$m_{\rm H}$~cm$^{-3})$ to resist expulsion by supernovae at all times.\label{fig:sex157mass}}
\end{figure}

The extent of mass loss occurring at perigalacticon is illustrated by the total mass contained within the central $3$~kpc versus time ($3$~kpc is a cut-off that captures the entire dwarf while ignoring most of any tidal streams produced).
The {ratio of mass, compared to the isolation model} versus time is shown for the two Sextans type dSphs above (perigalacticons of $63$ and $72$~kpc and apogalacticon of $150$~kpc) in Figs \ref{fig:sex156mass} and \ref{fig:sex157mass}.
As a dwarf passes by perigalacticon it first loses dark matter from within this $3$~kpc before then losing gas, although a much smaller fraction of gas is lost due to the more extended nature of the dark matter (for a Sextans type dwarf with a perigalacticon of $63$~kpc about $80\%$ of dark matter is lost at the first passage, while only $30\%$ of gas is lost).
Shortly after the dwarf passes the perigalacticon for the first time star formation shuts off within the dwarf, see \S\ref{ssec:SFR}, halting the gradual increase in stellar mass observed within the isolation model.

As the stellar population of dSphs is more centrally concentrated than that of the dark matter or gas -- a natural consequence of it only forming in denser gas -- it is much more resilient to tidal forces removing it from the dwarf, with the total stellar mass within the central three kpc remaining almost unchanged for the first two interactions, although some flattening of the profiles does occur (see Figures \ref{fig:sex156dens}--\ref{fig:sex157dens}).
This resilience is reflected in the fact that a dwarf loses a majority of its dark matter mass before it begins to lose stellar mass, decreasing the mass to light ratio by at least a factor of two in tidally stripped dSphs.

\begin{figure*}
  \includegraphics[width=\textwidth]{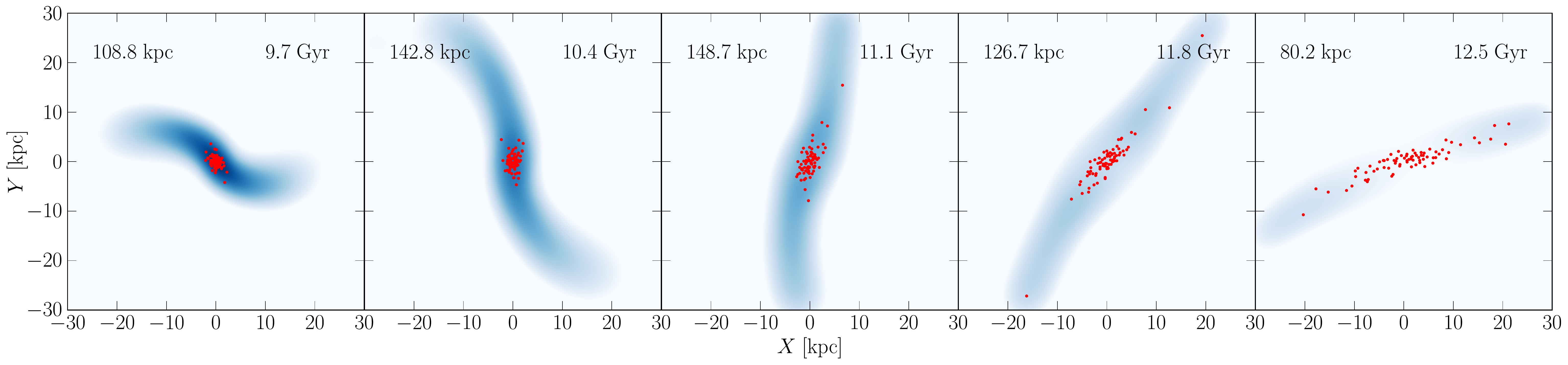}
  \caption{Projection of gas density inside the simulation of sex156-3 passages at a time of $9.7$--$12.5$~Gyr. The dwarf is currently located at a galactocentric radius of $149$~kpc, extremely close to its apogalacticon. The plot is centred on the mean position of the stellar particles. The position of every $50$th stellar particle is shown as a red point, the smoothing kernel is ignored for stellar particles for clarity. Here, gas that has been lowered in density by tidal stripping is able to be completely expelled by Type Ia supernova -- a process that begins as the dwarf approaches apogalacticon (panel $3$) -- forming a split tidal stream of gas.\label{fig:gasexpulsion}}
\end{figure*}

Tidal effects are also not the only cause of changes in the gas density profile for all dSphs.
For a subset of {orbits} (typically those with three interactions, an example is shown in Fig. \ref{fig:gasexpulsion}), tides remove a substantial portion of dark matter and gas from the dwarf galaxies, allowing Type Ia supernova to expel the remaining gas from the (now reduced) potential well of the dSph.
The limits of when this gas can be expelled is not entirely clear as tidal forces act in synergy with the supernovae and the removal of dark matter from the dwarf.
However, as far as a limit can be applied, the central gas density drops to $\sim$$3\times10^{-3}$~$m_{\rm H}$~cm$^{-3}$ in Sextans type dSphs before any bifurcation{, a sign of all gas being ejected from the dwarf,} is seen inside the gaseous stream { (for example, the right panel of Fig. \ref{fig:gasexpulsion})}.
In the more massive Sculptor dwarf this limit appears to be somewhat lower $\sim$$1\times10^{-3}$~$m_{\rm H}$~cm$^{-3}$, but is within the same ballpark.
The expulsion of gas, although the most obvious difference between the sex$156$ and sex$157$ simulations, happens simultaneously with the removal of stars and dark matter, and suggests that tidal remnants of dSphs will retain a lower fraction of gas than stars or dark matter.

\subsection{Star formation rate} \label{ssec:SFR}

\begin{figure}
  \includegraphics[width=0.5\textwidth]{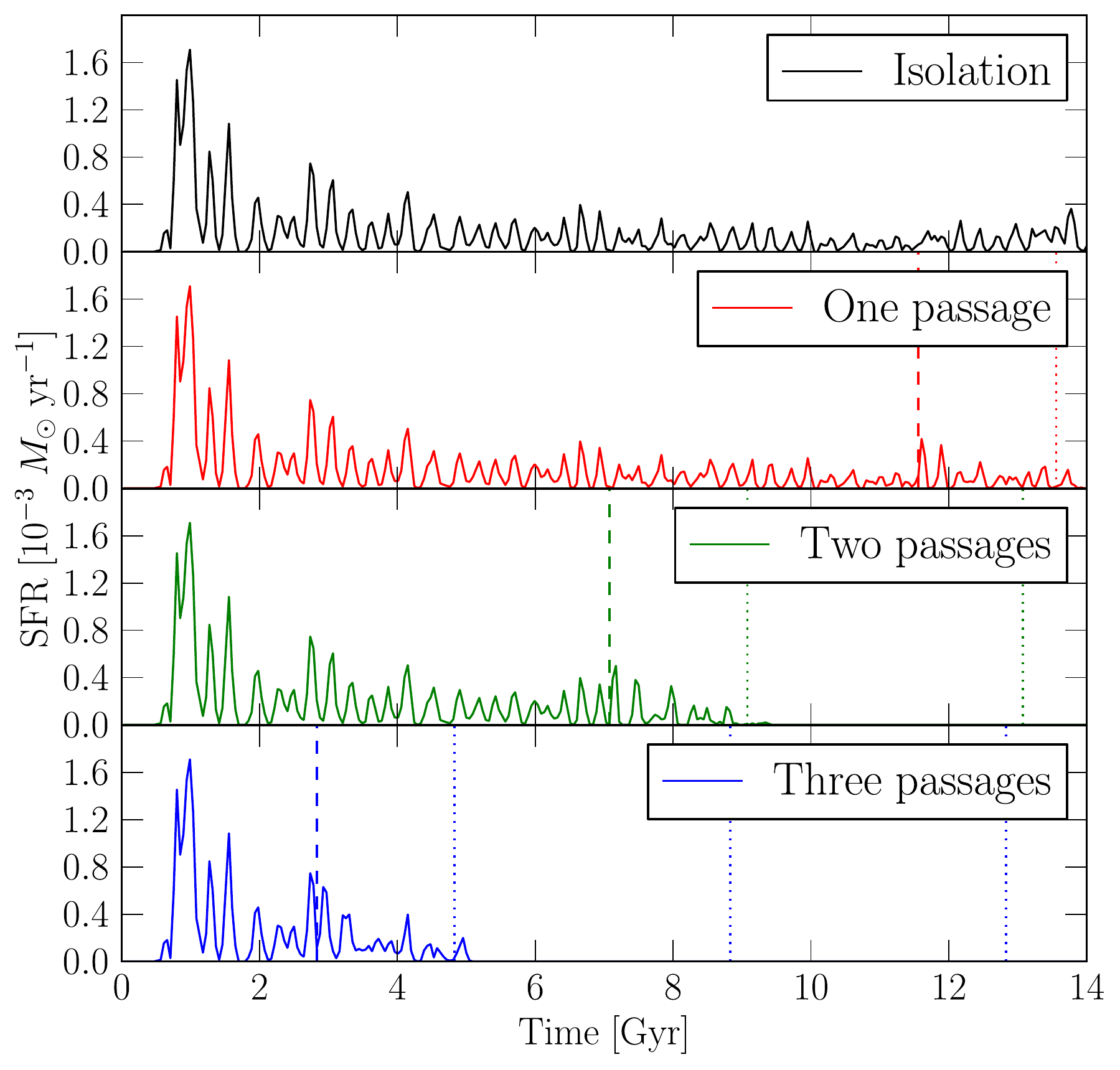} 
  \caption{Star formation history of the Sextans type dSph, sex156, shown in Fig. \ref{fig:sex156dens} compared with the Sextans model in isolation. The interaction of the dSph and the Milky Way begins at the dashed line for each model. Perigalacticons are indicated with a dotted line for each interaction model. The cessation of star formation clearly occurs around perigalacticon for models which experienced three and two perigalacticon passages respectively. The slight increase in star formation  observed at the start of the interaction is due to a numerical issue. Such an increase in star formation however, accounts for less than $5\%$ of the cumulative star formation inside a dwarf galaxy.\label{fig:sex156SFR}}
\end{figure}

The relationship between tidal forces and a dSphs star formation history has been examined previously \citep[e.g.][]{Brosch2004,Pasetto2011,Besla2012}.
One suggested effect is that tidally induced shocks may trigger a burst of star formation as a dwarf passes through perigalacticon \citep{Pasetto2011}.
Such an effect is however not seen inside the model dSphs with the perigalacticon shutting off star formation inside the dSphs as seen in Fig. \ref{fig:sex156SFR}.
In the following we try to understand the origin of this quenching by exploring peculiar cases which suffer lesser degrees of quenching.
\footnote{A slight increase in star formation (always comprising less than $5\%$ of the total stellar mass) is observed due to numerical issues.
Following \citet{Revaz2012}, the radiative cooling acting on a gas particle is shut off for a period of $5$~Myr post supernova feedback.
When starting a simulation from a snapshot (such as beginning the tidal interactions) the information containing the last feedback event (that occurs at an earlier time) is not available and the corresponding cooling shutoff is simply ignored.
In the absence of this adiabatic phase, the star formation is slightly enhanced.}

\begin{figure}
  \includegraphics[width=0.5\textwidth]{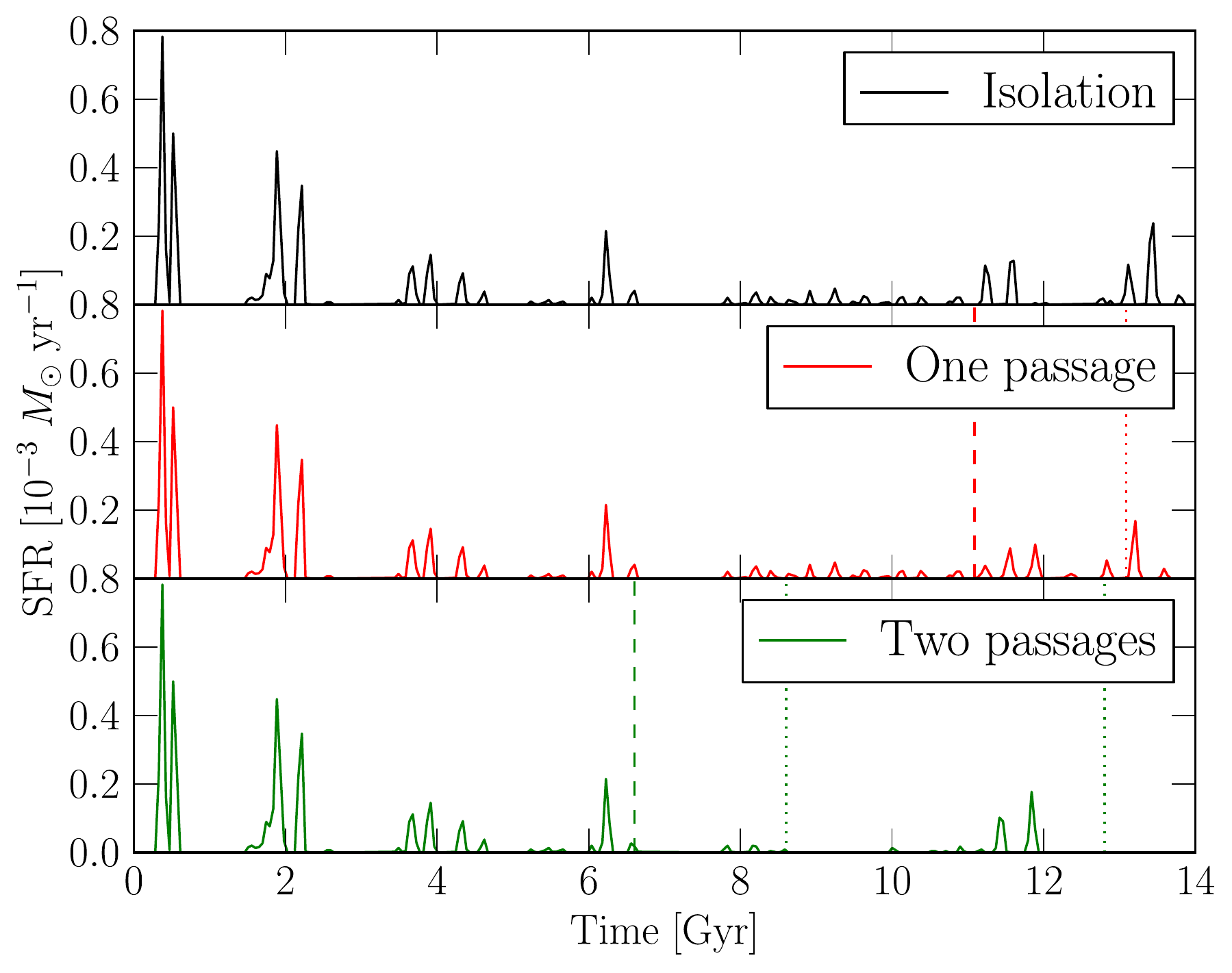} 
  \caption{Star formation history of a Carina type dSph, car157, with a perigalacticon and apogalacticon of $72$~kpc and $150$~kpc respectively compared to the Carina model in isolation. While other dwarf galaxies at this orbit experience a cessation of star formation at perigalacticon (similar to that seen in Fig. \ref{fig:sex156SFR}), the Carina dSph can continue to experience a few bursts of star formation at similar times to the isolated model. Such continued star formation occurs when there has not been any significant star formation at the perigalacticon of the orbit.\label{fig:car157SFR}}
\end{figure}

\begin{figure}
  \includegraphics[width=0.5\textwidth]{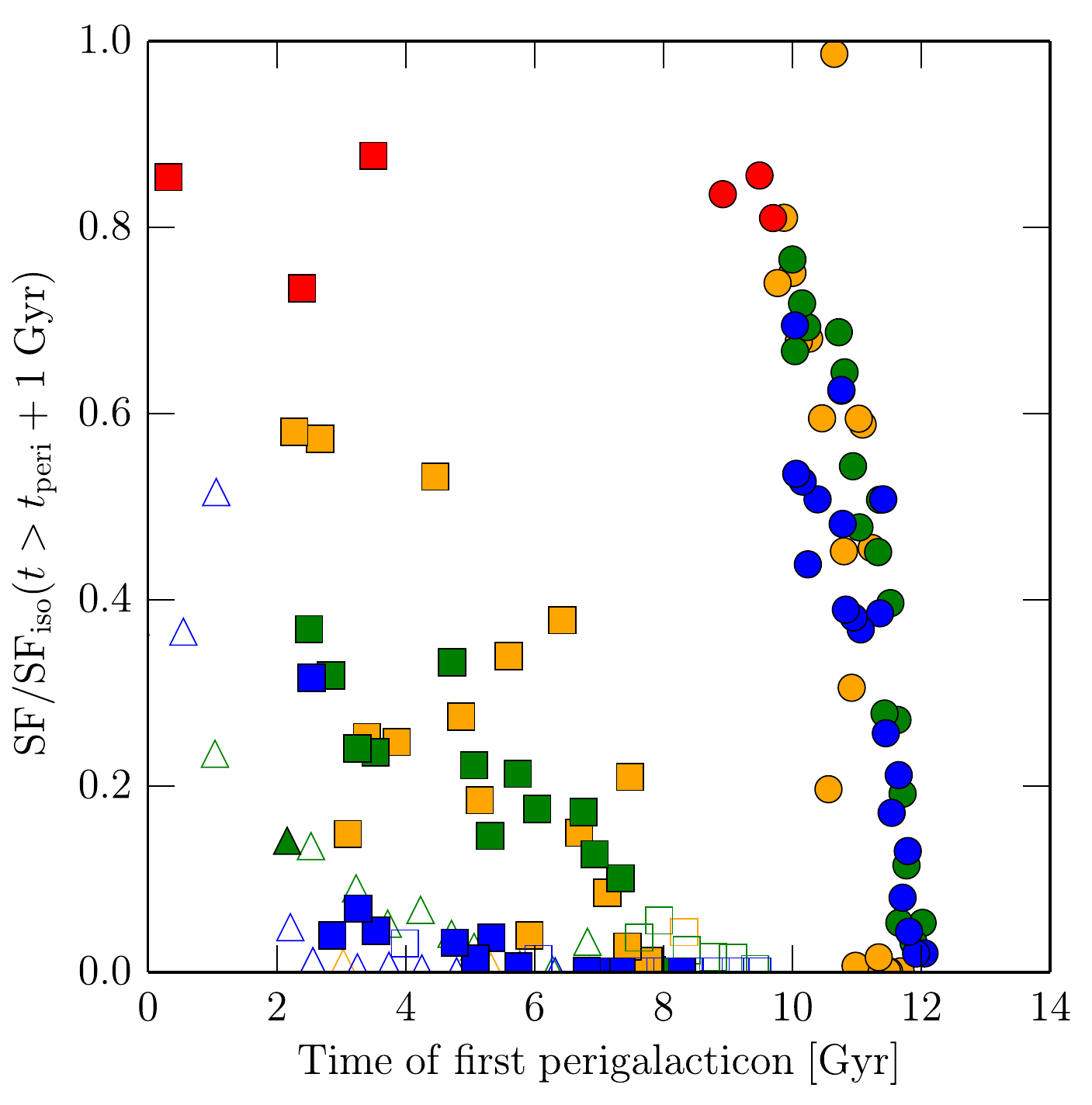}
  \caption{{Ratio of star formation that takes place after $1$~Gyr post perigalacticon compared to the same time in the isolation models.
Carina type dSphs are shown in orange, Sextans type green, Sculptor type blue and Fornax type red. Solid symbols represent those dSphs which retain over $80\%$ of their stellar mass from the central $3$~kpc, while open symbols represent dSphs which have lost over $20\%$ of their stellar mass.
The dSphs which have only undergone one perigalacticon passage are shown as circles, two passages squares and three passages as triangles.
Fornax, located further out than the other model dSphs shows the greatest resilience to quenching.
Of the other model dSphs (which share the same orbits) there is a slight trend of decreasing resilience to quenching with increasing mass.
Particularly for models which approached perigalacticon at an early time.}}
\label{fig:SFRratio}
\end{figure}

Low mass dwarf models illustrate a greater degree of resilience to the disruptive effects of the tidal forces than other dSphs at any given orbit, as shown in Fig. \ref{fig:car157SFR} for model car157 {and Fig. \ref{fig:SFRratio}.
Here, particularly for two passage models, the lower mass Carina dwarf tends to have more star formation (relative to the isolation model) than Sextans, which itself has more than Sculptor. Fornax like models, which have perigalacticons further out have even more star formation, although whether this is due to its mass or orbit is not clear.}

Notably, Carina type models only survive quenching if they are in a quiescent phase at perigalacticon.
In this situation, the lack of any supernova near perigalacticon (and consequently lack of outflowing gas) prevents tidal forces from stripping large quantities of gas.
Some Sculptor type dwarf models, all with a larger perigalacticon of $72$~kpc, are able to restart star formation but only after an extended period of no star formation in contrast to the isolation model.
Similarly Fornax type models, which all have higher perigalacticons compared to other models are able to restart star formation but only after a quiescent phase.
Such a correlation between supernovae and the quenching -- temporary or otherwise -- of dSphs, suggests that even medium to high perigalacticons, the synergy between the tidal forces moving gas outwards and remnant supernova keeps the density of the remaining gas too low to form stars.
Sufficiently massive dSphs, more massive than any models here, will be able to keep the gas sufficiently dense and quickly reaccrete outflowing gas to prevent this synergistic quenching from occurring even temporarily.

That result that tidal forces quench star formation after one passage corresponds well with the radial distribution of dSph, which can be produced through tidal interactions if and only if one perigalacticon passage is sufficient to transform a dwarf into a dSph \citep{Slater2013}.

\subsection{Metallicity}

Much of the impact of tidal interactions on the metallicity will be due to the star formation shut-off inside the dwarf galaxy.
As the star formation shuts-off around perigalacticon, tidal stirring of gas is unable to mix the remaining gas and is therefore unlikely to impact the metallicity of stars within the dSph.
A secondary factor is that tidal stripping disproportionately strips stars at (comparatively) large radii, where high metallicity stars are unlikely to form.

Observationally we are unable to resolve the majority of stars in a dSph and are limited to the brightest stars, typically those in the red giant branch (RGB) phase of stellar evolution.
As the proportion of stars which are in the RGB phase changes with age, we weight the metallicity of each stellar particle -- representing a cluster of stars -- in the simulation by the proportion of stars which {are climbing the RGB.}
{Using the {\em BaSTI} stellar tracks \citep{Pietrinferni2013} we consider the metallicity and age of each stellar particle, and interpolate to calculate the mass of stars which would have left the main sequence, but have not yet reached the tip of the RGB.
Using the chosen IMF \citep{Kroupa2001}, we then calculate what number of stars this represents and weight each particle accordingly.}

\begin{figure*}
  \includegraphics[width=\textwidth]{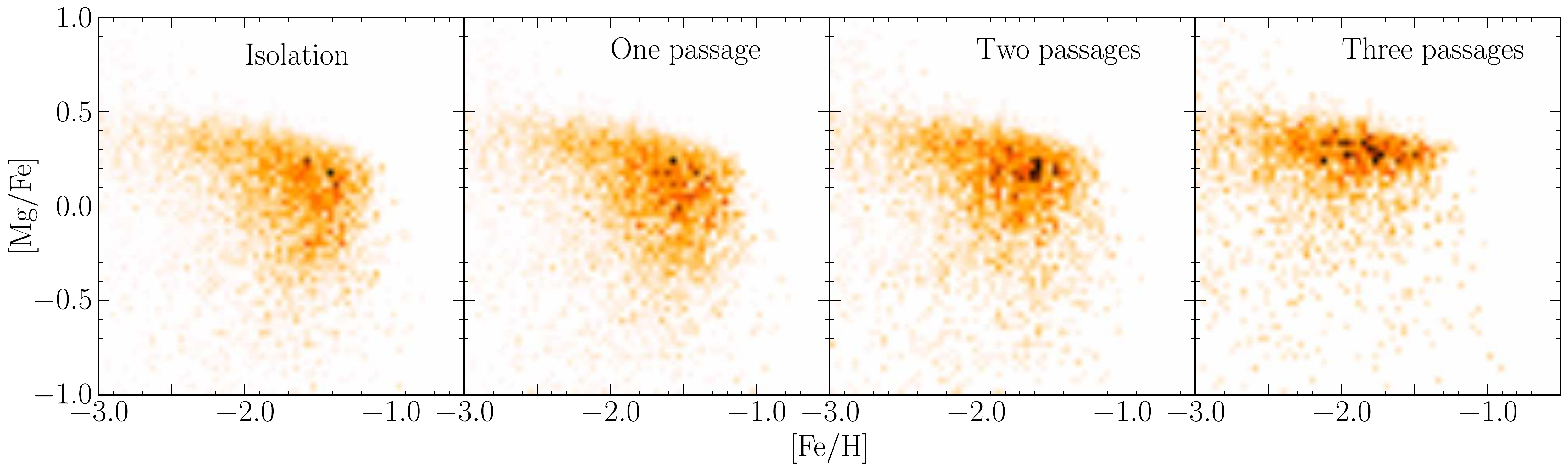}
  \caption{Metallicity of RGB stars of model sex156 a Sextans type dSph with a perigalacticon and apogalacticon of $63$~kpc and $150$~kpc respectively.
    As perigalacticon passages shut down the star formation within a dSph (\S\ref{ssec:SFR}) models which have had multiple interactions will form fewer stars from Type Ia Supernova enriched gas and subsequently contain fewer [$\alpha$/Fe] poor stars at high [Fe/H].
    Observationally such an effect may be obscured if the gas in the dSph is poorly mixed, such as in the Carina dSph.\label{fig:sex156MgFe}}
\end{figure*}

\begin{figure}
  \includegraphics[width=0.5\textwidth]{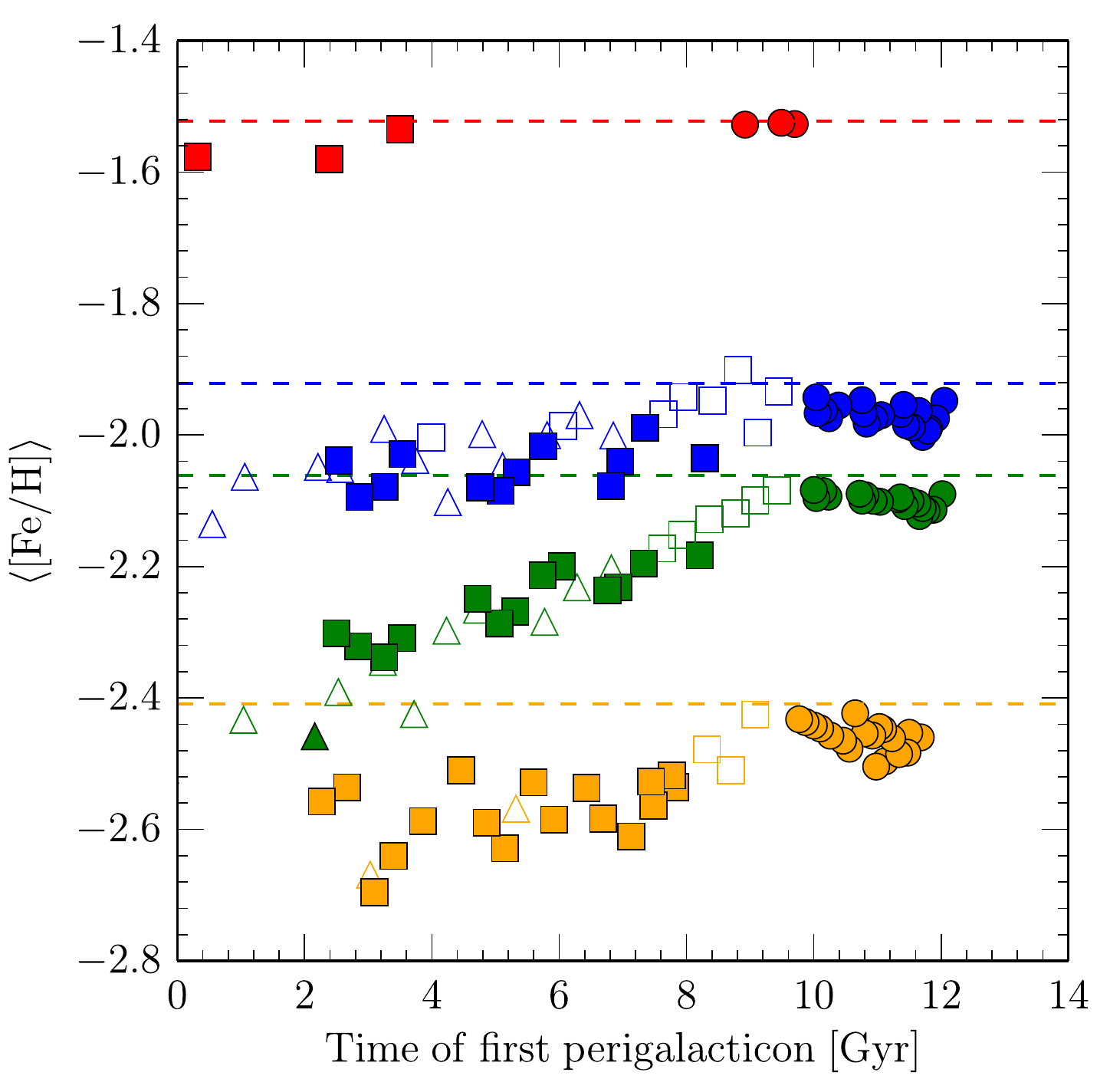}
  \caption{Mean metallicity, $\langle$[Fe/H]$\rangle$, of RGB stars within a $3$~kpc radius of model dSph versus  time of first perigalacticon.
    From bottom to top groups of simulation points correspond to Carina type dSphs (orange), Sextans (green), Sculptor (blue) and Fornax (red).
    The dSphs which retain over $80\%$ of their stars are solid {symbols}, while open {symbols} represent dSphs which retain less.
    {The dSphs which have only undergone one perigalacticon passage are shown as circles, two passages squares and three passages as triangles.}
    The mean metallicity of each isolation model is shown as a dashed line in the relevant colour.
    \label{fig:FeHvT}}
\end{figure}

As the number of passages determines when the star formation will shut off the effects on metallicity show a progression from the least affected at one passage to the most massively affected at three passages, illustrated in Fig. \ref{fig:sex156MgFe} for model sex156.
Nearly all this variation in metallicity seems to occur due to star formation being shut off at perigalacticon.
That this variation arises from the difference in shut off time (and hence length of star formation) is illustrated in Fig. \ref{fig:FeHvT}, where the mean metallicity versus time of first perigalacticon passage is shown for all simulations.

\begin{figure*}
  \includegraphics[width=\textwidth]{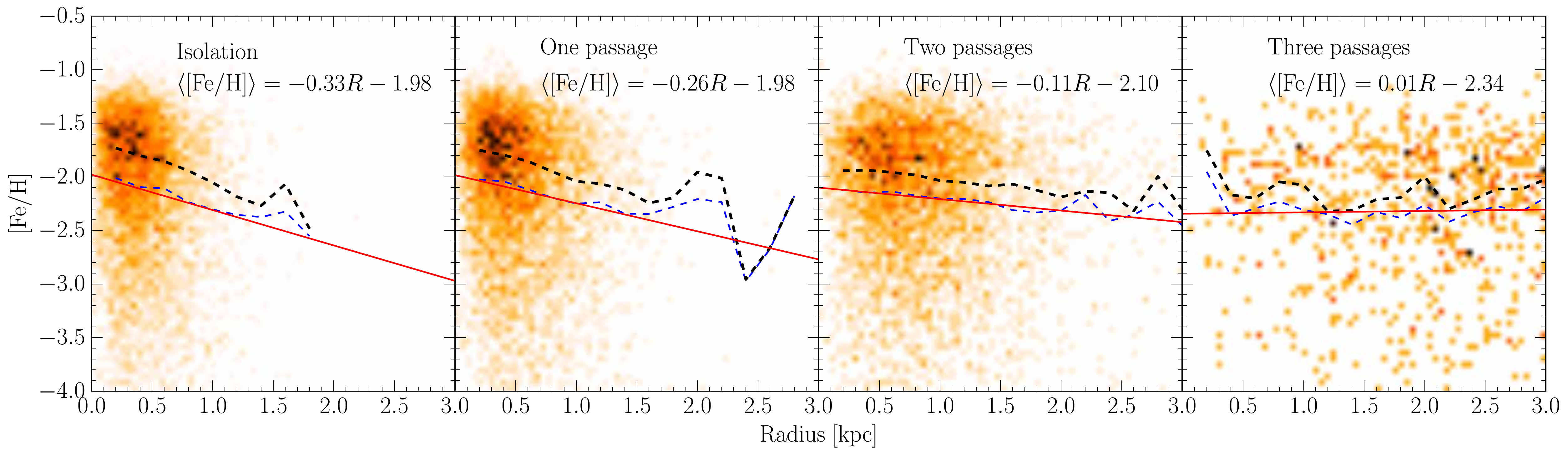}
  \caption{[Fe/H] versus radius of RGB stars inside model sex156, a Sextans type dSph with a perigalacticon of $63$~kpc and an apogalacticon of $150$~kpc (model sex156).
    The colour from light yellow to dark red shows the relative amount of points within each [Fe/H]-R bin, normalized to the highest density within each individual model.
    The thin dashed (blue) line shows the mean metallicity ($\langle$[Fe/H]$\rangle$) within each radial $200$~pc bin from the centre, the thick dashed (black) line shows the median metallicity within each bin while the solid (red) line shows the least squares linear regression of the metallicity versus radius (at the distance of Sextans, $1$~kpc = $40'$).
    Isolation (and one passage) models show a clear drop in mean metallicity with increasing radial distance from the centre.
    The decrease in the gradient arises from the lack of young stars in the centre of dwarf models which have undergone multiple passages (as the tidal forces help quench the dwarf) and tidal forces smoothing out the stellar density profile(see Fig. \ref{fig:sex156dens} and Fig. \ref{fig:sex156-gradZold}).\label{fig:sex156-gradZ}}
\end{figure*}

\begin{figure*}
  \includegraphics[width=\textwidth]{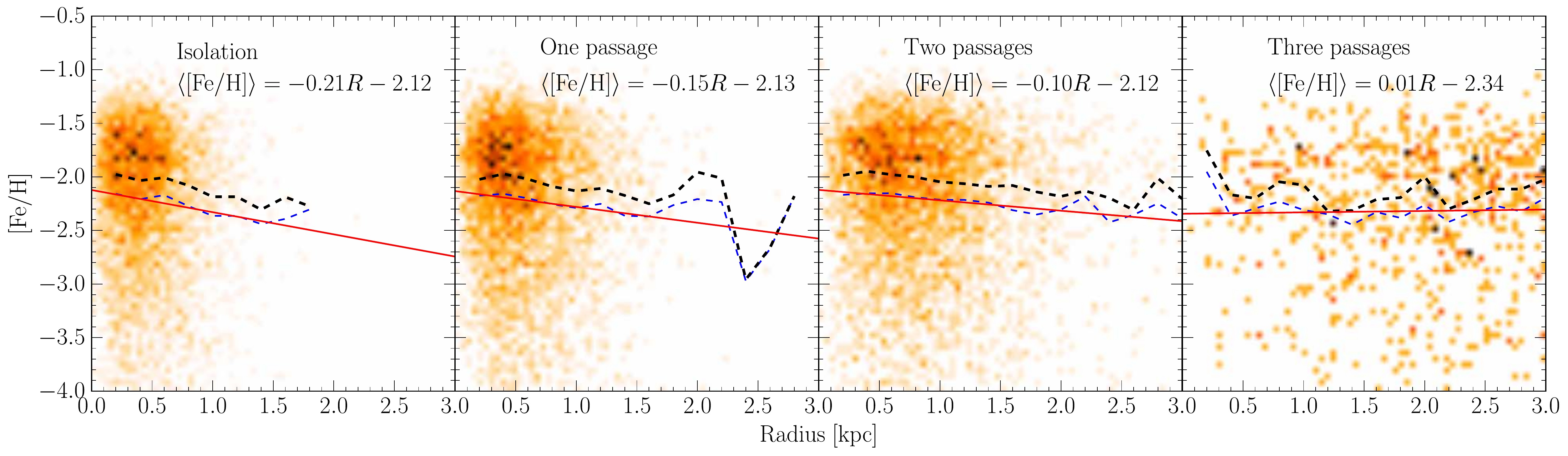}
   \caption{Same as Fig. \ref{fig:sex156-gradZ} but only accounting for old stars ($>6$~Gyr).
     The scaling and line types are as in Fig. \ref{fig:sex156-gradZ}.
     Compared to Fig. \ref{fig:sex156-gradZ}, here only old stars are selected, such an age cutoff is chosen to represent the stellar population that existed before the quenching of star formation in the two passage model (sex156-2).
     Hence the Isolation, One passage and Two passage panels display the same stellar populations (the three passage model [sex156-3] was quenched before this cutoff, however, with its massive disruption any gradient would be destroyed anyway).
     The age cutoff hence allows an estimation of the effect of tidal forces in smoothing any metallicity gradient present within the dwarf models.
     Restricting the considered stellar population to only these old stars a lower gradient is seen due to a smoothing of the stellar density profile, indicating that tidal forces may lower the metallicity gradient present within dSphs even if no stars are tidally removed.
    \label{fig:sex156-gradZold}}
\end{figure*}

Even though dSphs manage to maintain most of their stars (\S\ref{ssec:dens}), the tidal interactions do sufficiently perturb the dSphs to gradually erase any metallicity gradient present within the dSphs with increasing number of interactions as shown in Fig. \ref{fig:sex156-gradZ}.
Such an effect is seen, although weaker, even when considering only old stars formed before most interactions have taken place as seen in Fig. \ref{fig:sex156-gradZold}.
Compared to \citet{Revaz2012} { where star formation was artificially truncated}, the gradient in the isolation case is much stronger, {as young, relatively metal rich, stars had the opportunity to form in the centre.}

\begin{figure}
  \includegraphics[width=0.5\textwidth]{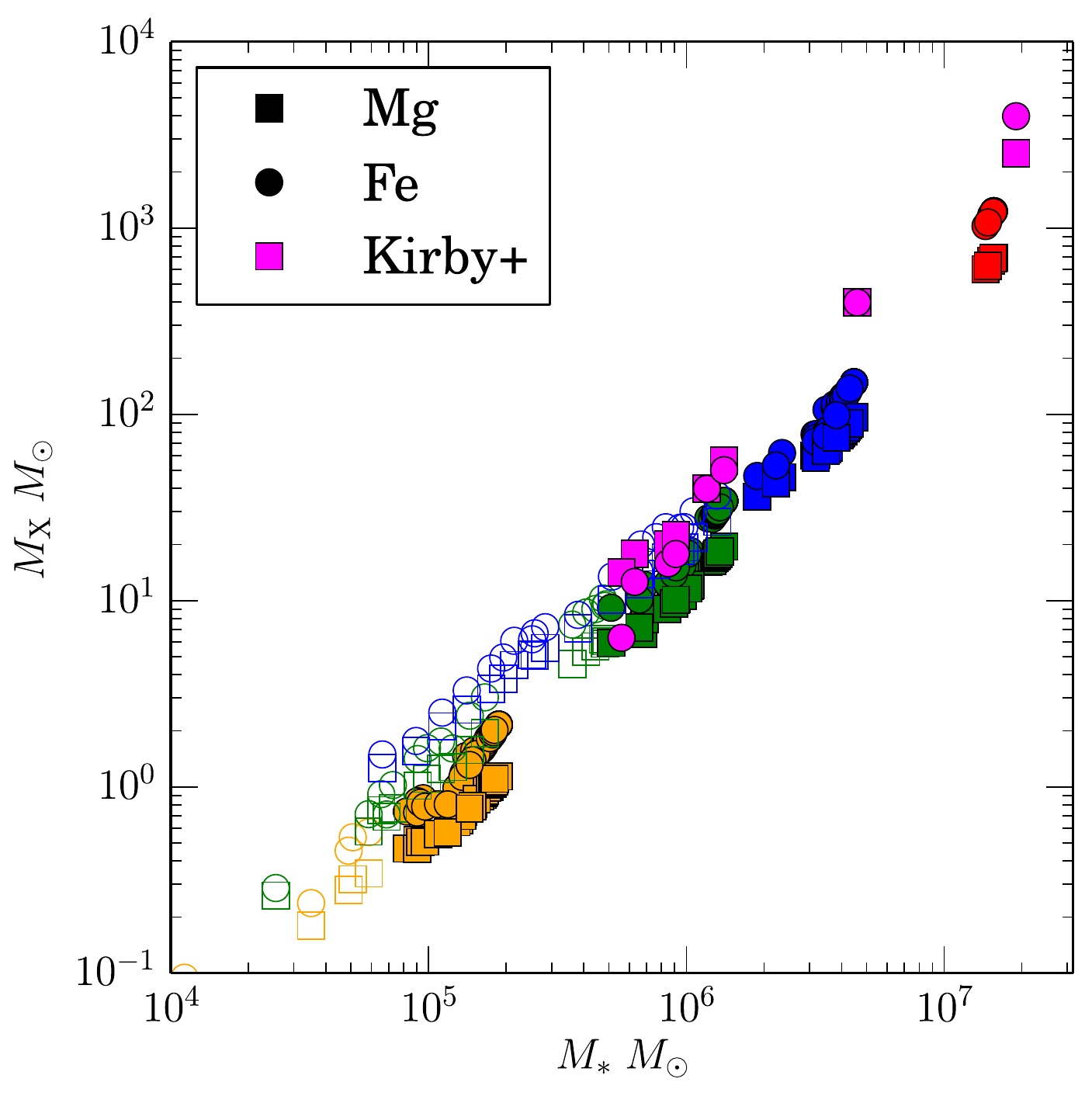}
  \caption{The mass of retained metals within $3$~kpc as a function of stellar mass within the same radius. Models which retain less than $80\%$ of their stellar mass are shown as {open} symbols. {For clarity, all dSphs, regardless of how many perigalacticons are shown as either circles for Fe or squares for Mg}. dSphs which suffer substantial mass loss retain more metals than smaller undisturbed systems, a sign of higher metallicity in the core. In undisturbed cases, the model dSphs contain less metals (by a factor of $\sim3$) than the observed dSphs at the same stellar mass {a possible discrepancy that arises from differing IMFs and yields}. The left to right groups of simulation points correspond to Carina type dSphs (orange), Sextans (green), Sculptor (blue) and Fornax (red). Observed values for Mg and Fe from \citet{Kirby2011b} are shown in magenta. \label{fig:Zmass}}
\end{figure}

Recently \citet{Kirby2011} presented observations and calculations of the total metal content inside some Local Group dwarf galaxies.
Intriguingly, they find that the total metal content is much lower than the expected metal production by supernovae throughout the dSphs life time.
{\citet{Kirby2011b} suggest that this discrepancy, between expected metal production and the stellar abundances, should be due to galactic winds.}
We examine our simulations to test if such an expulsion of metals is required to produce such a low fraction of ``retained'' metals inside the stars, or whether metals are generally inefficient at being transferred from supernovae ejecta to new stars.
In Fig. \ref{fig:Zmass} we display the total mass of Mg and Fe contained in stellar particles as a function of stellar mass of the dwarf.
Our undisturbed models, those which lose less than $20\%$ of their stars, under produce metals -- compared to the \citeauthor{Kirby2011b} sample -- for stellar masses above $\sim$$10^6$~$M_\sun$, although, like real dSphs, the model dSphs are inefficient at reprocessing metals.

{Despite this inefficiency,} more massive dSphs are able to recycle metals more efficiently than less massive dSphs.
Subsequently, dSphs which then lose a substantial portion  $($$\gta$$20\%$$)$ possess a higher metal mass than dSphs of the same (final) stellar mass which never lost stars.
These stripped dSphs however, possess a lower relative difference between the total mass of Mg and Fe, suggesting that tidally striped dSphs will have a smaller variance in $[$Mg$/$Fe$]$ vs $[$Fe$/$H$]$ than dSphs that have evolved without these tidal interactions.
Fornax is the most efficient dwarf at reprocessing matter -- unsurprising as it is the most massive -- however, it still manages to reprocess less than $10\%$ of the produced metals in stars, with the other dwarf models reprocessing $1$--$5\%$ of the produced metals, comparable to the observed Local Group dSphs \citep{Kirby2011b}.

Metals which are produced in these supernovae are retained in the surrounding gas, which may be removed from the dwarf at any time, such as through ram pressure stripping.
The presence of galactic winds, may be responsible for the removal of this gas and along with it, the metals, however, most stars will have been formed before these metals are produced, and the amount of metals recycled into stars will likely always be low inside dSphs.

\begin{figure}
  \includegraphics[width=0.5\textwidth]{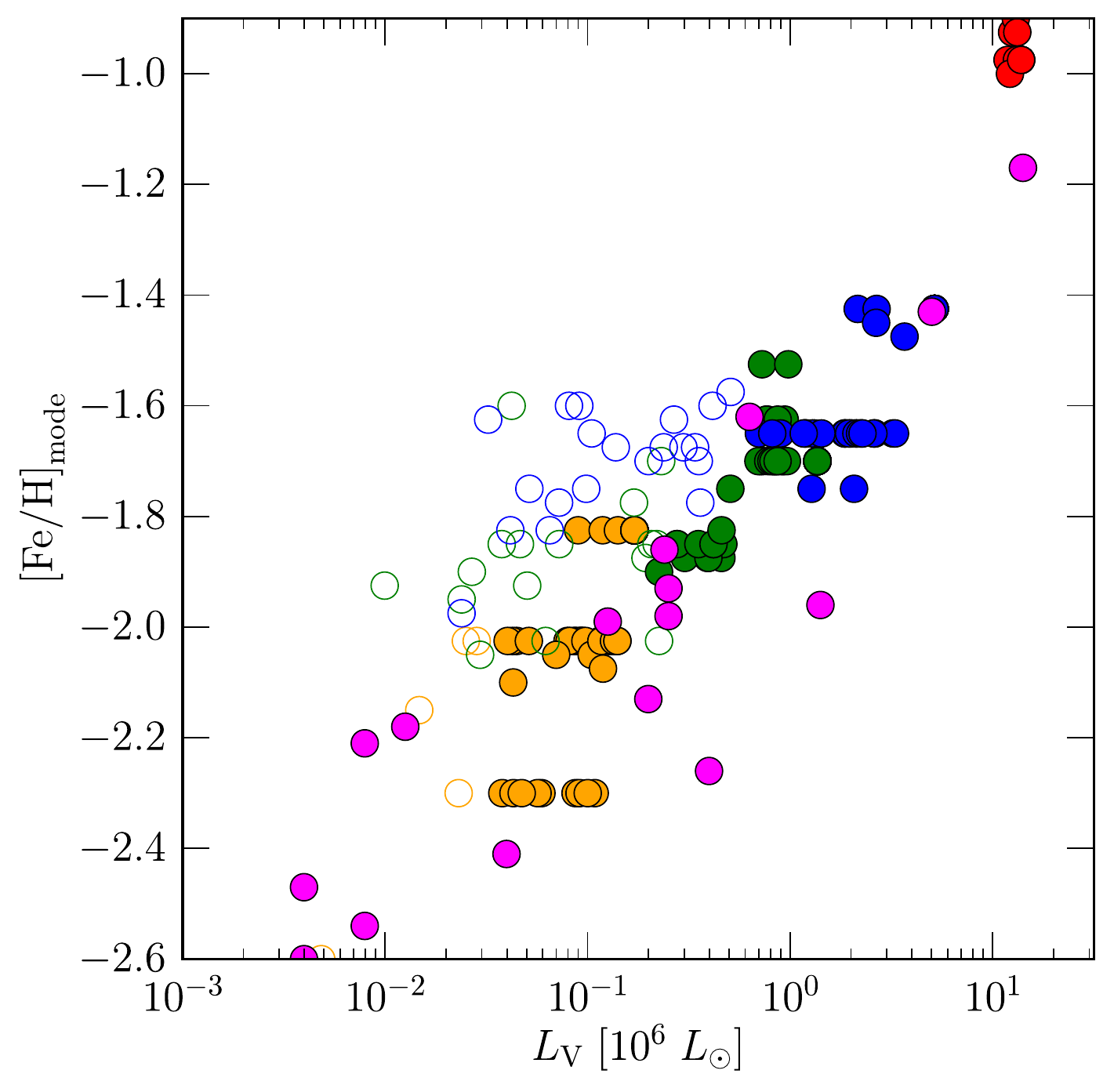}
  \caption{The mode of the metallicity distribution of dSphs within $3$~kpc of their centre as a function of luminosity within the same radius. Models which retain less than $80\%$ of their stellar mass are shown as {open circles. For clarity all dSphs are shown as circles, regardless of the number of perigalacticons}. dSphs which suffer substantial mass loss show a higher mean metallicity than smaller undisturbed systems. In undisturbed cases, the model dSphs have a lower $\langle$[Fe/H]$\rangle$ (by up to $0.5$~dex) than the observed dSphs at the same luminosity. The left to right groups of simulation points correspond to Carina type dSphs (orange), Sextans (green), Sculptor (blue) and Fornax (red). Observational values are shown in magenta.\label{fig:ZLum}}
\end{figure}

The effect is also seen if one compares the luminosity of the dwarf to the mean metallicity, shown in Fig. \ref{fig:ZLum}.
The observed luminosities for the four classical dSphs considered are from \citep{Walker2009} and metallicity distributions from \citet{Helmi2006}, \citet{Battaglia2011}, \citet{Tolstoy2004} and \citet{Battaglia2006} for Carina, Sextans, Sculptor and Fornax respectively; luminosities and (mean) metallicity for the remaining, mostly, ultra-faint dSphs are from \citet{Kirby2011c}.
The use of mean metallicity does not greatly affect the observed values, which are often well represented by only slightly skewed Gaussian curves.
Here, tidally disrupted dSphs tend to move {to the left} as they retain their mode metallicity but drop in luminosity.
That disrupted dSphs are able to move {to the left} along this relation suggests that at least some of the Local Group dSphs may have undergone tidal stripping in their past.
{Such a process would help} to explain their present day metallicity, particularly at low luminosity.
We note that even if a dwarf loses $80\%$ of its initial stellar mass, a core of dark matter and stars often remains which although extended could be recognised as a dSph.

\subsection{Effects of resolution}
{In order to cover the wide parameter space here the models were necessarily of a medium resolution.
However, to ensure that the resolution of the models adequately represent the dynamical evolution of the dSphs, we have undertaken a few higher resolution models as listed in Tables \ref{tab:dSphparam}.
In the high resolution cases however, the mass of the stellar particles becomes very small, and can not really be considered as a full stellar system.
As we considered a continuous IMF \citep[as in][]{Revaz2012} rather than a discretely sampled IMF this problem is partially counteracted.

The present day density of high resolution models of an isolation version of the Sextans dSph models and a two passage model of sex156 \footnote{We focus on this model as it was the most used throughout the paper, although higher resolution runs were undertaken in isolation for all dwarfs and in interaction for car156 and all showed substantially similar results.} (that is a perigalacticon of $63$~kpc and apogalacticon of $150$~kpc) are shown in Fig. \ref{fig:hires dens}.
In isolation the models converge well, with minimal differences.
When the models have undergone two tidal interactions however, the inner kiloparsec retains slightly more material, in particular retaining slightly more gas.
Once again the star formation is quenched, as can be seen by a decrease in the central density decrease compared to the isolation models and in the star formation history shown in Fig. \ref{fig:hires-sfr}.

\begin{figure}
\includegraphics[width=0.5\textwidth]{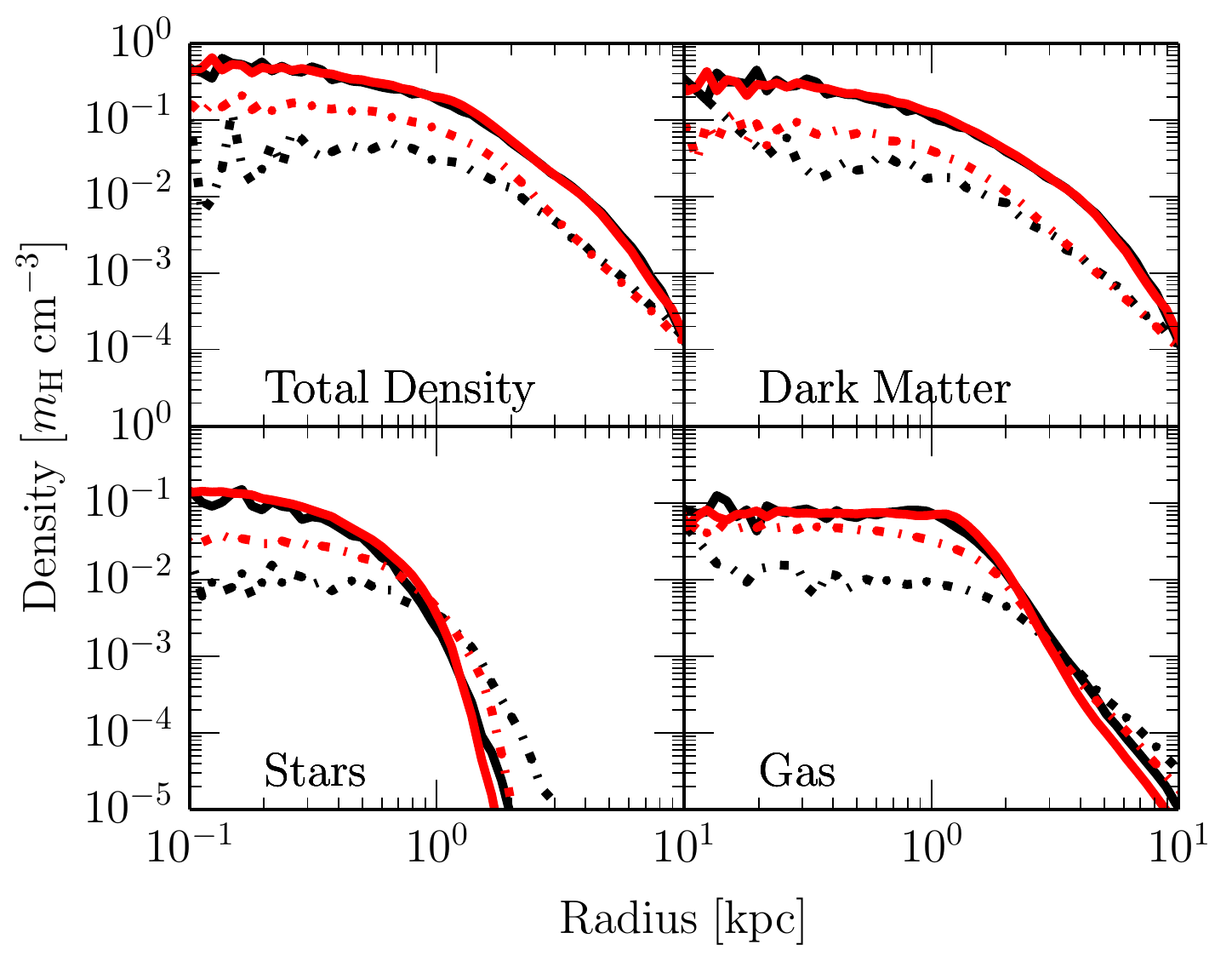}
\caption{{Density of a Sextans like dwarf in isolation both in a normal resolution (black) and high resolution (red) case and after two interactions at a perigalacticon of $63$~kpc and apogalacticon of $150$~kpc.
The isolated case is shown as a solid line and the interaction case a dot-dash line for both.}}\label{fig:hires dens}
\end{figure}

\begin{figure}
\includegraphics[width=0.5\textwidth]{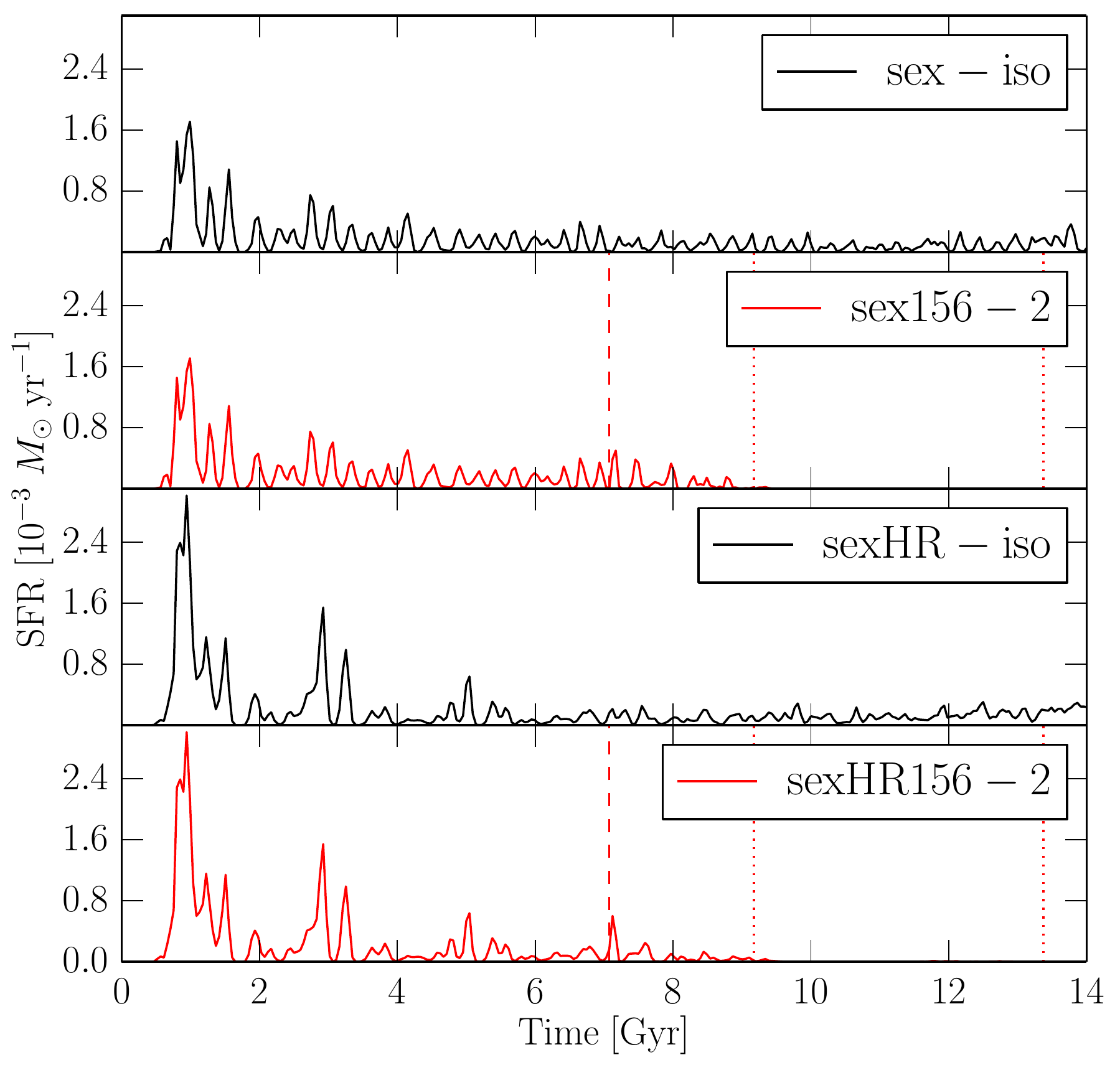}
\caption{{Star formation of a Sextans like dwarf in isolation and undergoing two passages at a perigalacticon of $63$~kpc and an apogalacticon of $150$~kpc.
The top two panels in show the generic resolution model (sex-iso in black, sex156-2 in red respectively) and the bottom two the high resolution models (sexHR-iso in black, sexHR156-2 in red respectively).
The beginning of the interaction is shown with a dashed line and the time of perigalacticons shown with a dotted line.
Here the star formation is quenched in both cases soon after the perigalacticon passage.}}\label{fig:hires-sfr}
\end{figure}

Despite the slightly higher star formation, the metallicity of the dwarfs is unaffected by the increase as seen in Fig. \ref{fig:hiresFe} and Fig. \ref{fig:hiresintFe}.
In both cases the interactions prevent stars reaching the low $\alpha/$Fe characteristic of Type Ia supernova suggesting that, at least qualitatively, the results of the medium resolution runs are sound.

\begin{figure}
  \includegraphics[width=0.5\textwidth]{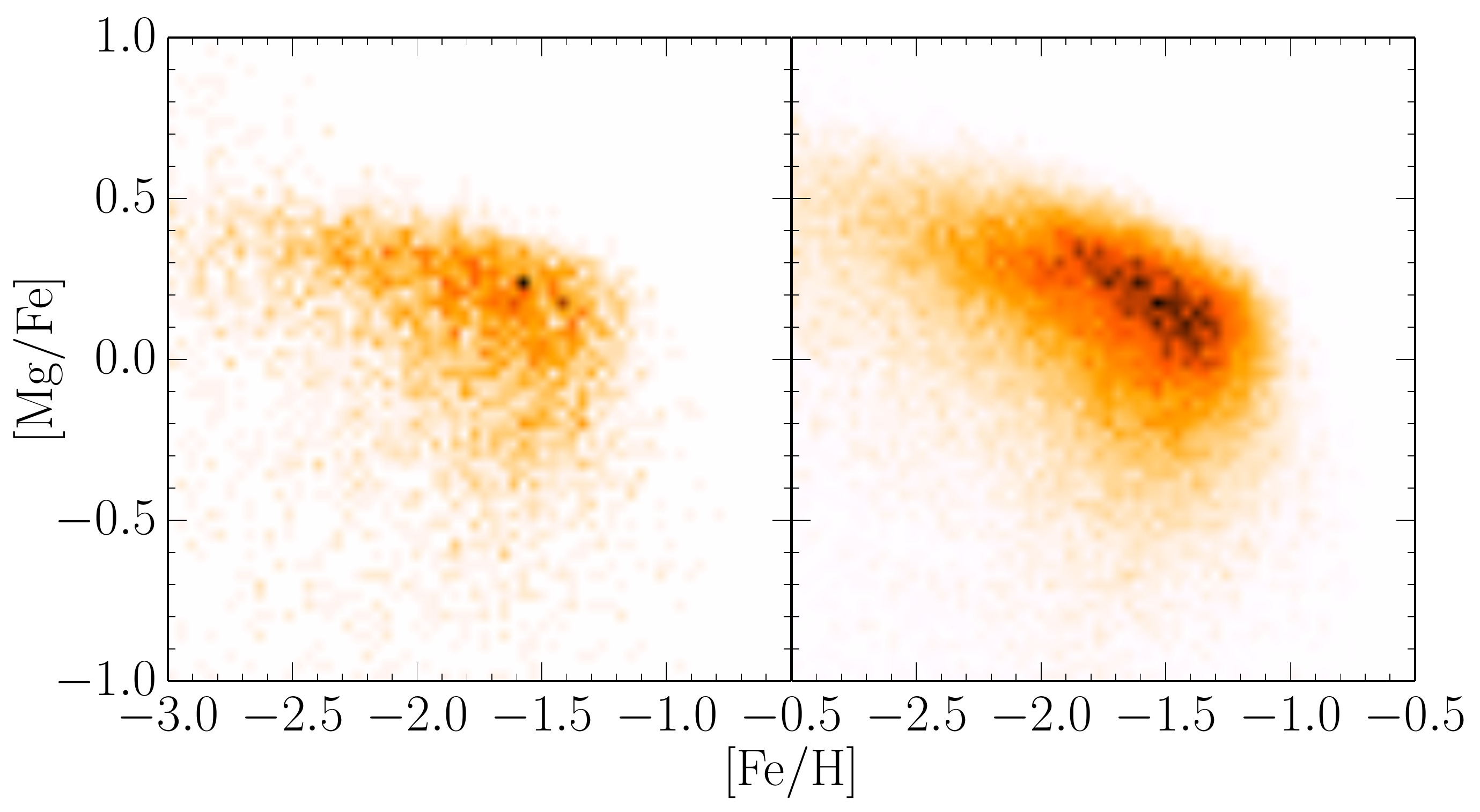}
  \caption{{[Mg/Fe] versus [Fe/H] for a Sextans type dSph in isolation at the generic resolution (left, model sex-iso) and high resolution (right, model sexHR-iso).}}
  \label{fig:hiresFe}
\end{figure}

\begin{figure}
  \includegraphics[width=0.5\textwidth]{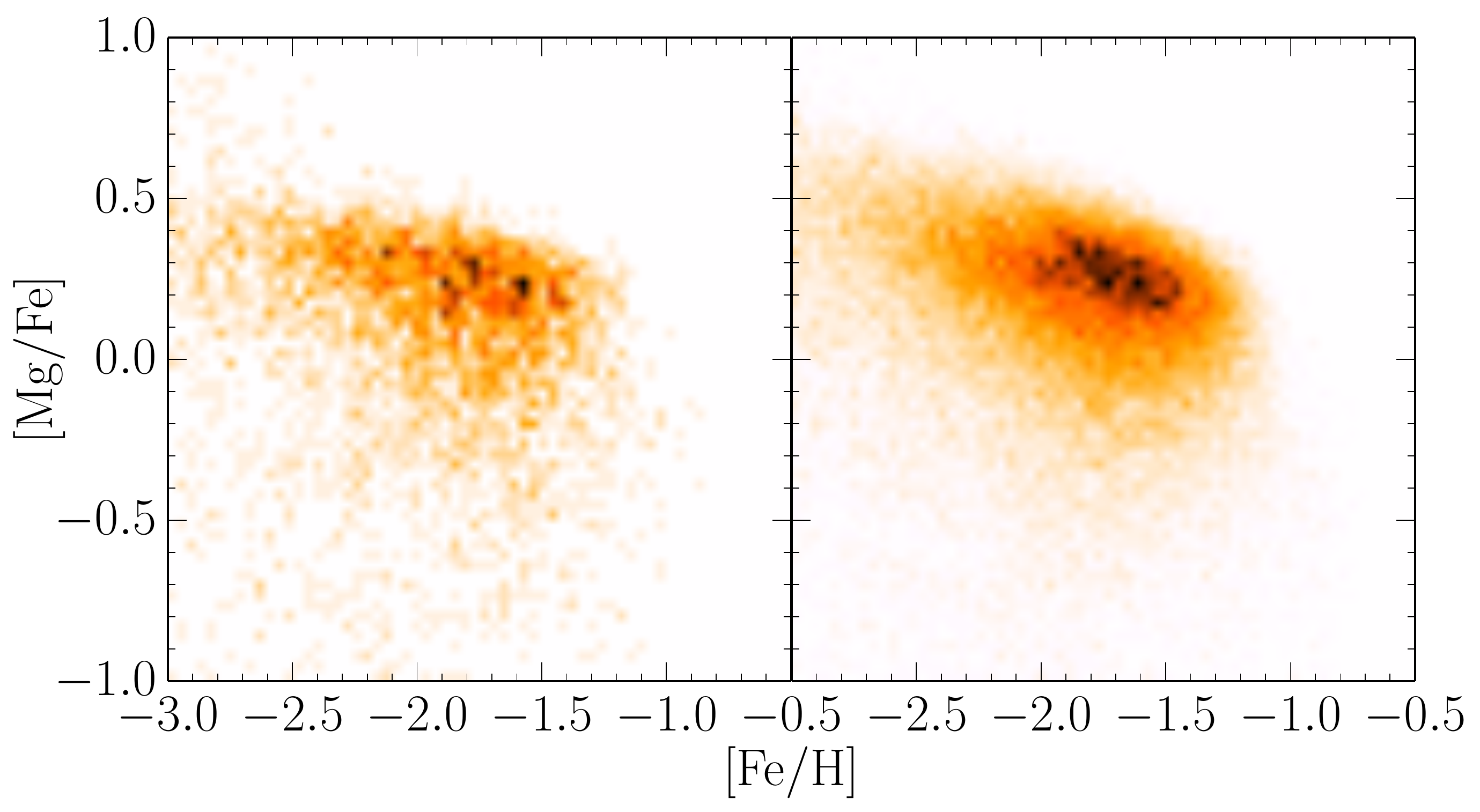}
  \caption{{ [Mg/Fe] versus [Fe/H] for a Sextans type dSph that has undergone two orbits with a apogalacticon of $150$~kpc and a perigalacticon of $63$~kpc  at generic resolution (left, model sex156-2) and high resolution (right, model sexHR156-2).}}
  \label{fig:hiresintFe}
\end{figure}

}

\section{Discussion and Conclusion}\label{sec:conclusion}

We have undertaken a series of simulations to examine how the chemodynamical properties of dwarf spheroidals are altered through tidal interactions.
That major chemodynamical changes, excluding those arising from star formation quench, were only minor, although potentially noticeable, in undisrupted dSphs suggests that the isolation models provide a good base to understanding how dSphs have evolved through cosmic time.

As our base models, which was evolved in isolation were chosen to closely match the observed properties of dSphs when evolved in isolation \citep{Revaz2012}, na\"\i{}ve comparisons with observations will bias towards those models with the smallest tidal impacts. 
However, as nearly all dSphs that experienced drastic changes in observable properties such as metallicity or luminosity (see also Appendix B) became noticeably disrupted, such a bias is likely to be small when considering at least the classical dSphs.

{The quenching of star formation seems to occur in more massive dwarfs, and biased towards those undergoing star formation near the perigalacticon.
If the stellar feedback is weaker than proposed, such an effect may be reduced to the outer edges of a dwarf, or disappear entirely.
As this effect was somewhat mass dependent, the dwarfs most likely to experience such a quenching near perigalacticon are those which underwent star formation before infall and hence are able to blow gas away from the central star forming region.
}

{
Due to the large parameter space that satellite galaxies inhabit and computational restraints there are limitations to the modelling able to be undertaken.
The choice of a cored profile arises from the optimisation of the best fit models in isolation, with the initial conditions for dSphs in an NFW profile so far unknown.
Replacing the density profile, without changing any other properties, does allow a dwarf slightly more resilience to tidal stripping.
But such a dwarf also over produces stars for the same baryon/dark-matter ratio, and consequently a different chemical history, suggesting that a lower mass NFW profile may be appropriate.
In this case, the tidal effects still lead to disruption and for the masses considered, a quenching of star formation, however, the central region may maintain more consistency than the cored profile examined here.}

Our simulations were also undertaken with a mass of the Milky Way halo on the low tail of the distribution proposed in the literature \citep{Boylan-Kolchin2011,Kafle2012} although still in good agreement with observations.
Using a more massive halo is likely to extend its sphere of influence to disrupt dSphs even further out than that proposed here.
{Such a result would be in contrast to that if run inside a growing cosmological simulation. Due to the lower mass at early times, the tidal forces could be expected to be lower, however, as the evolution of orbits within a growing halo are heavily phase dependent it is difficult to say definitively.}

Our investigation also ignored the ram pressure effects of a hot halo, such a hot halo is likely to act in synergy with tidal stripping to remove the low density gas from the dSphs potential, allowing the dwarf to then expel any gas that survives (as shown in Fig. \ref{fig:gasexpulsion} but occurring over a much larger orbital range).
We also explored only a small portion of the parameter space of dwarf progenitors, more massive dSphs may be expected to better survive tidal interactions while least massive will be more easily disrupted.

We find that\begin{itemize}
\item tidal forces during one perigalacticon passage are sufficient to quench star formation even in gas endowed dSphs,
\item the metallicity gradient is affected by the number of passages, with more tidal interactions resulting in a lower metallicity gradient,
\item dSphs are generally inefficient at the recycling of metal from supernovae and will in all cases only have a few percent of metals recycled no matter the state of remaining gas.
\end{itemize}
The quenching of star formation is due to synergy between supernovae and tidal forces keeping gas at sufficiently low densities to prevent star formation and we do not see any gas compression in these dSphs which may allow further star formation to occur during these interactions.
Such quenching suppresses the latest generation of stars, an effect which imprints into the $\alpha$/Fe abundance ratios within the dSphs.
The halting of star formation supports the suggestion that the radial distribution of dSphs arises from tidal interactions \citep{Slater2013}.

A portion, approximately half, of the lower metallicity gradient was also due to the quenched star formation, however, tidal interactions do partially erase metallicity gradients through purely tidal effects.

Such tidal interactions also shift dSphs to a higher $\langle$[Fe/H]$\rangle$/L ratio, allowing a greater range of the observed distribution covered than dSphs that have only evolved in isolation, albeit only when$\gta$$20\%$ of stars are lost.
The dSphs which undergo two passages are able to lose $30\%$ of their initial gas mass purely to tidal forces and internal processes, while which only undergo one passage can only lose $5\%$ the same way, with dSphs that lose more tending to undergo too much disruption to be recognisable as classical dSphs today.
Such small gas losses, when nearly all dSphs are \ion{H}{i} deficient, indicates that other processes are necessary to strip the majority of a dSphs initial gas.
Ram pressure is the most likely process for this stripping, with it able to remove the gas from a dwarf rapidly \citep[e.g.][]{Gatto2013}.
Dwarf winds may also contribute, although such winds are not needed to instantly remove metals from the dSphs and would have to work in conjunction with ram pressure to produce the observed radial distribution of gas-rich/gas-deficient dwarfs \citep[see][]{Grcevich2009}.

\begin{acknowledgements} The authors gratefully acknowledge support from the Swiss National Science Foundation. The work greatly benefited from the International Space Science Institute (ISSI) in Bern, thanks to the funding of the team “First Stars in dwarf galaxies”. Simulation outputs were processed with the parallelised Python package, \texttt{pNbody} (\texttt{http://lastro.epfl.ch/projects/pNbody}).
\end{acknowledgements}

\FloatBarrier
\appendix{

\section{Initial Parameters}
\begin{table}
  \caption{Carina type}
 \centering\begin{tabular}{lccccc}
\hline\hline
{Run}&{Apo}&{Peri}&{$T_{\rm iso,1}$}&{$T_{\rm iso,2}$}&{$T_{\rm iso,3}$}\\ &{[kpc]}& {[kpc]} & {[Gyr]}& {[Gyr]}& {[Gyr]}\\
  \hline
  car103	&	$105$	&	$35.8$	&	$11.7$	&	$9.1$	&	$6.5$\\
  car104	&	$105$	&	$44.6$	&	$11.5$	&	$8.7$	&	$5.9$\\
  car105	&	$105$	&	$53.3$	&	$11.3$	&	$8.3$	&	$5.3$\\
  car106	&	$105$	&	$61.9$	&	$11.1$	&	$7.9$	&	$4.6$\\
  car107	&	$105$	&	$70.5$	&	$11.0$	&	$7.5$	&	$4.0$\\
  car153	&	$150$	&	$36.2$	&	$11.5$	&	$7.8$	&	$4.1$\\
  car154	&	$150$	&	$45.2$	&	$11.3$	&	$7.4$	&	$3.5$\\
  car155	&	$150$	&	$54.1$	&	$11.2$	&	$7.1$	&	$2.9$\\
  car156	&	$150$	&	$62.9$	&	$11.1$	&	$6.7$	&	$2.3$\\
  car157	&	$150$	&	$71.7$	&	$11.0$	&	$6.4$	&	$1.8$\\
  car203	&	$200$	&	$36.5$	&	$10.9$	&	$5.9$	&	$0.9$\\
  car204	&	$200$	&	$45.5$	&	$10.8$	&	$5.5$	&	$0.3$\\
  car205	&	$200$	&	$54.5$	&	$10.7$	&	$5.2$	&	$-$\\
  car206	&	$200$	&	$63.5$	&	$10.5$	&	$4.8$	&	$-$\\
  car207	&	$200$	&	$72.5$	&	$10.5$	&	$4.4$	&	$-$\\
  car253	&	$250$	&	$36.6$	&	$10.2$	&	$3.8$	&	$-$\\
  car254	&	$250$	&	$45.7$	&	$10.1$	&	$3.4$	&	$-$\\
  car255	&	$250$	&	$54.8$	&	$10.0$	&	$3.0$	&	$-$\\
  car256	&	$250$	&	$63.9$	&	$9.9$	&	$2.7$	&	$-$\\
  car257	&	$250$	&	$73.0$	&	$9.8$	&	$2.3$	&	$-$\\
  \hline\end{tabular}
\label{tabA:paramCar}
  \tablefoot{Simulation parameters for Carina type dSphs. $T_{{\rm iso},n}$ is the time the dSph evolves in isolation before passing the perigalacticon $n$ times. Apo is the Apogalacticon and Peri is the perigalacticon.}

\end{table}

\begin{table}
  \caption{Sextans type}
  \centering\begin{tabular}{lccccc}
\hline\hline{Run}&{Apo}&{Peri}&{$T_{\rm iso,1}$}&{$T_{\rm iso,2}$}&{$T_{\rm iso,3}$}\\ &{[kpc]}& {[kpc]} & {[Gyr]}& {[Gyr]}& {[Gyr]}\\
  \hline
  sex103	&	$105$	&	$35.8$	&	$12.0$	&	$9.4$	&	$6.8$\\
  sex104	&	$105$	&	$44.6$	&	$11.9$	&	$9.1$	&	$6.3$\\
  sex105	&	$105$	&	$53.3$	&	$11.8$	&	$8.7$	&	$5.7$\\
  sex106	&	$105$	&	$61.9$	&	$11.7$	&	$8.4$	&	$5.2$\\
  sex107	&	$105$	&	$70.5$	&	$11.7$	&	$8.2$	&	$4.7$\\
  sex153	&	$150$	&	$36.2$	&	$11.6$	&	$8.0$	&	$4.3$\\
  sex154	&	$150$	&	$45.2$	&	$11.5$	&	$7.6$	&	$3.7$\\
  sex155	&	$150$	&	$54.1$	&	$11.4$	&	$7.3$	&	$3.1$\\
  sex156	&	$150$	&	$62.9$	&	$11.3$	&	$7.0$	&	$2.6$\\
  sex157	&	$150$	&	$71.7$	&	$11.3$	&	$6.7$	&	$2.1$\\
  sex203	&	$200$	&	$36.5$	&	$11.0$	&	$6.0$	&	$1.0$\\
  sex204	&	$200$	&	$45.5$	&	$10.9$	&	$5.7$	&	$0.4$\\
  sex205	&	$200$	&	$54.5$	&	$10.8$	&	$5.3$	&	$-$\\
  sex206	&	$200$	&	$63.5$	&	$10.7$	&	$5.0$	&	$-$\\
  sex207	&	$200$	&	$72.5$	&	$10.7$	&	$4.7$	&	$-$\\
  sex253	&	$250$	&	$36.6$	&	$10.4$	&	$3.9$	&	$-$\\
  sex254	&	$250$	&	$45.7$	&	$10.2$	&	$3.5$	&	$-$\\
  sex255	&	$250$	&	$54.8$	&	$10.1$	&	$3.2$	&	$-$\\
  sex256	&	$250$	&	$63.9$	&	$10.0$	&	$2.8$	&	$-$\\
  sex257	&	$250$	&	$73.0$	&	$10.0$	&	$2.5$	&	$-$\\
  \hline\end{tabular}
\label{tabA:paramSex}
\tablefoot{Simulation parameters for Sextans type dSphs. $T_{{\rm iso},n}$ is the time the dSph evolves in isolation before passing the perigalacticon $n$ times.}
\end{table}

\begin{table}
  \caption{Sculptor type}
  
\centering\begin{tabular}{lccccc}\hline\hline
{Run}&{Apo}&{Peri}&{$T_{\rm iso,1}$}&{$T_{\rm iso,2}$}&{$T_{\rm iso,3}$}\\ &{[kpc]}& {[kpc]} & {[Gyr]}& {[Gyr]}& {[Gyr]}\\
  \hline
  scl103	&	$105$	&	$35.8$	&	$12.1$	&	$9.5$	&	$6.9$\\
  scl104	&	$105$	&	$44.6$	&	$11.9$	&	$9.1$	&	$6.3$\\
  scl105	&	$105$	&	$53.3$	&	$11.8$	&	$8.8$	&	$5.8$\\
  scl106	&	$105$	&	$61.9$	&	$11.7$	&	$8.5$	&	$5.2$\\
  scl107	&	$105$	&	$70.5$	&	$11.8$	&	$8.3$	&	$4.8$\\
  scl153	&	$150$	&	$36.2$	&	$11.7$	&	$8.0$	&	$4.3$\\
  scl154	&	$150$	&	$45.2$	&	$11.5$	&	$7.6$	&	$3.7$\\
  scl155	&	$150$	&	$54.1$	&	$11.4$	&	$7.3$	&	$3.2$\\
  scl156	&	$150$	&	$62.9$	&	$11.4$	&	$7.0$	&	$2.6$\\
  scl157	&	$150$	&	$71.7$	&	$11.4$	&	$6.8$	&	$2.1$\\
  scl203	&	$200$	&	$36.5$	&	$11.1$	&	$6.0$	&	$1.0$\\
  scl204	&	$200$	&	$45.5$	&	$10.9$	&	$5.7$	&	$0.4$\\
  scl205	&	$200$	&	$54.5$	&	$10.8$	&	$5.3$	&	$-$\\
  scl206	&	$200$	&	$63.5$	&	$10.8$	&	$5.0$	&	$-$\\
  scl207	&	$200$	&	$72.5$	&	$10.8$	&	$4.7$	&	$-$\\
  scl253	&	$250$	&	$36.6$	&	$10.4$	&	$3.9$	&	$-$\\
  scl254	&	$250$	&	$45.7$	&	$10.2$	&	$3.6$	&	$-$\\
  scl255	&	$250$	&	$54.8$	&	$10.1$	&	$3.2$	&	$-$\\
  scl256	&	$250$	&	$63.9$	&	$10.1$	&	$2.8$	&	$-$\\
  scl257	&	$250$	&	$73.0$	&	$10.0$	&	$2.6$	&	$-$\\
  \hline\end{tabular}
\label{tabA:paramScl}

  \tablefoot{Simulation parameters for Sculptor like dSphs. $T_{{\rm iso},n}$ is the time the dSph evolves in isolation before passing the perigalacticon $n$ times.}
\end{table}

\begin{table}
  \caption{Fornax type}
  \centering\begin{tabular}{lccccc}\hline\hline{Run}&{Apo}&{Peri}&{$T_{\rm iso,1}$}&{$T_{\rm iso,2}$}&{$T_{\rm iso,3}$}\\ &{[kpc]}& {[kpc]} & {[Gyr]}& {[Gyr]}& {[Gyr]}\\
  \hline
  fnx151	&	$150$	&	$106.4$	&	$9.5$	&	$3.8$	&	$-$\\
  fnx171	&	$170$	&	$107.1$	&	$9.7$	&	$3.4$	&	$-$\\
  fnx201	&	$200$	&	$107.9$	&	$9.5$	&	$2.4$	&	$-$\\
  fnx251	&	$250$	&	$108.9$	&	$8.9$	&	$0.3$	&	$-$\\
  \hline\end{tabular}
\label{tabA:paramFnx}
  \tablefoot{Simulation parameters for Fornax like dSphs. $T_{{\rm iso},n}$ is the time the dSph evolves in isolation before passing the perigalacticon $n$ times.}
  
\end{table}

\section{Summary of properties}
\begin{table*}
\caption{Carina~type}\centering\begin{tabular}{lcccccccc}\hline\hline{Model}&{$L_V$~[$10^{5}$~$L_\sun$]}&{$r_{1/2}$~[kpc]}&{$\langle$[Fe/H]$\rangle$$_{r_{1/2},{\rm RGB}}$}&{$\sigma_{v,{r_{1/2}}}$~[km~s$^{-1}$]}&{SFR$_{99}$~Gyr}&{$r_{S}$~[kpc]}&{$M_{\star}$(3~kpc)/$M_\star$}&{$M_{\rm g}(3$~kpc$)/M_{\rm g}$}\\
\hline
car-iso	&    1.66	& 0.33 &-2.13 & 4.27 &13.44 & 0.65 & 1.00 & 0.73 \\
car103-1	&   0.819	& 1.15		&-2.23		& 1.36		&11.79		& 4.06		& 0.88		& 0.14		\\
car103-2	&   0.359	& 2.43		&-2.43		& 1.40		& 9.20		&14.07		& 0.38		& 0.02		\\
car104-1	&   0.866	& 1.12		&-2.19		& 2.27		&11.60		& 3.68		& 0.93		& 0.14		\\
car104-2	&    0.33	& 2.47		&-2.36		& 1.88		& 8.73		&11.40		& 0.35		& 0.02		\\
car105-1	&   0.898	& 0.66		&-2.17		& 2.08		&11.60		& 1.27		& 1.00		& 0.30		\\
car105-2	&    0.38	& 2.21		&-2.40		& 1.24		& 8.86		&10.94		& 0.41		& 0.02		\\
car106-1	&   0.856	& 0.43		&-2.17		& 3.00		&11.57		& 0.82		& 1.00		& 0.53		\\
car106-2	&   0.537	& 1.92		&-2.39		& 1.49		& 8.41		& 6.55		& 0.68		& 0.06		\\
car107-1	&   0.834	& 0.36		&-2.18		& 3.44		&11.55		& 0.67		& 1.00		& 0.63		\\
car107-2	&    0.83	& 0.41		&-2.36		& 3.19		&13.24		& 0.87		& 1.00		& 0.56		\\
car153-1	&   0.882	& 0.77		&-2.17		& 2.06		&11.60		& 1.63		& 1.00		& 0.30		\\
car153-2	&   0.543	& 2.19		&-2.42		& 1.45		& 8.39		& 8.92		& 0.65		& 0.01		\\
car154-1	&   0.891	& 0.53		&-2.16		& 2.69		&11.60		& 1.13		& 1.00		& 0.46		\\
car154-2	&   0.467	& 1.95		&-2.42		& 1.63		& 7.55		& 7.15		& 0.61		& 0.07		\\
car155-1	&    1.15	& 0.47		&-2.08		& 2.48		&13.08		& 1.01		& 1.00		& 0.53		\\
car155-2	&   0.557	& 0.92		&-2.36		& 1.79		& 7.31		& 1.93		& 0.99		& 0.23		\\
car156-1	&    1.25	& 0.42		&-2.11		& 2.85		&13.16		& 0.84		& 1.00		& 0.59		\\
car156-2	&   0.871	& 0.44		&-2.32		& 3.27		& 6.84		& 0.76		& 1.00		& 0.59		\\
car157-1	&    1.25	& 0.42		&-2.17		& 2.83		&13.20		& 0.72		& 1.00		& 0.65		\\
car157-2	&   0.773	& 0.42		&-2.23		& 3.07		&11.85		& 0.78		& 1.00		& 0.56		\\
car203-1	&    1.11	& 0.52		&-2.12		& 2.07		&11.91		& 1.08		& 1.00		& 0.40		\\
car203-2	&   0.491	& 1.94		&-2.40		& 1.18		& 6.26		& 4.94		& 0.71		& 0.06		\\
car204-1	&    1.13	& 0.48		&-2.12		& 2.43		&11.65		& 1.03		& 1.00		& 0.48		\\
car204-2	&   0.527	& 1.94		&-2.41		& 1.28		& 7.93		& 5.97		& 0.68		& 0.05		\\
car205-1	&    1.37	& 0.48		&-2.07		& 2.43		&13.05		& 0.96		& 1.00		& 0.54		\\
car205-2	&   0.525	& 0.71		&-2.39		& 2.28		& 6.61		& 1.30		& 1.00		& 0.38		\\
car206-1	&   0.934	& 0.37		&-2.13		& 3.54		&11.59		& 0.73		& 1.00		& 0.65		\\
car206-2	&   0.659	& 0.41		&-2.24		& 3.05		& 6.39		& 0.76		& 1.00		& 0.58		\\
car207-1	&    1.01	& 0.35		&-2.13		& 2.97		&11.65		& 0.69		& 1.00		& 0.68		\\
car207-2	&    1.04	& 0.37		&-2.07		& 2.70		&12.90		& 0.90		& 1.00		& 0.57		\\
car253-1	&    1.13	& 0.53		&-2.10		& 2.11		&12.69		& 1.08		& 1.00		& 0.42		\\
car253-2	&    0.51	& 1.97		&-2.44		& 1.33		& 6.40		& 4.50		& 0.75		& 0.05		\\
car254-1	&    1.33	& 0.56		&-2.12		& 2.16		&13.09		& 1.05		& 1.00		& 0.46		\\
car254-2	&   0.494	& 1.38		&-2.47		& 2.11		& 5.26		& 4.38		& 0.87		& 0.10		\\
car255-1	&    1.31	& 0.45		&-2.12		& 2.52		&13.20		& 0.85		& 1.00		& 0.59		\\
car255-2	&    0.47	& 0.58		&-2.50		& 2.81		& 4.11		& 0.98		& 1.00		& 0.52		\\
car256-1	&    1.36	& 0.41		&-2.09		& 3.29		&12.96		& 0.80		& 1.00		& 0.61		\\
car256-2	&    1.15	& 0.50		&-2.07		& 2.59		&12.98		& 0.91		& 1.00		& 0.46		\\
car257-1	&    1.36	& 0.37		&-2.11		& 2.78		&13.01		& 0.75		& 1.00		& 0.66		\\
car257-2	&    0.96	& 0.35		&-2.19		& 3.37		&11.51		& 0.79		& 1.00		& 0.65		\\
\hline\end{tabular}
\label{tab:carres}
  \tablefoot{Properties at the present day for a Carina type dSph in the present day. The dash after the simulation name indicates how many perigalacticon passages the dwarf has experienced, with iso indicating the isolation model. Columns from left to right are $L_V$, the V-band luminosity; $r_{1/2}$, the half-light radius; $\langle$[Fe/H]$\rangle$$_{r_{1/2},{\rm RGB}}$, the mean [Fe/H] of RGB stars within the half-light radius; $\sigma_{v,{r_{1/2}}}$, the velocity dispersion at the half-light radius; SFR$_{99}$, the time at which $99\%$ of star formation has taken place, an indication of the quenching time; $r_S$, the Sersic scale radius; $M_{\star}$(3~kpc)/$M_\star$, the ratio of stellar mass within $3$~kpc of the dSph centre to all stars formed in the corresponding simulation, including before ``infall''; and $M_{\rm g}(3$~kpc$)/M_{\rm g}$, the ratio of gaseous mass within the central $3$~kpc to all gas mass in the simulation.}
\end{table*}

\begin{table*}

\caption{Sextans~type}\centering\begin{tabular}{lcccccccc}\hline\hline{Model}&{$L_V$~[$10^{5}$~$L_\sun$]}&{$r_{1/2}$~[kpc]}&{$\langle$[Fe/H]$\rangle$$_{r_{1/2},{\rm RGB}}$}&{$\sigma_{v,{r_{1/2}}}$~[km~s$^{-1}$]}&{SFR$_{99}$~Gyr}&{$r_{S}$~[kpc]}&{$M_{\star}$(3~kpc)/$M_\star$}&{$M_{\rm g}(3$~kpc$)/M_{\rm g}$}\\
\hline
sex-iso	&    13.5	& 0.49 &-1.84 & 5.04 &13.61 & 0.73 & 1.00 & 0.78 \\
sex103-1	&    7.32	& 0.98		&-1.85		& 2.87		&12.27		& 3.38		& 0.95		& 0.21		\\
sex103-2	&     3.5	& 2.59		&-2.06		& 1.23		& 9.67		&13.76		& 0.41		& 0.01		\\
sex104-1	&    6.99	& 0.90		&-1.85		& 2.99		&12.03		& 2.61		& 0.99		& 0.24		\\
sex104-2	&    3.21	& 2.61		&-2.05		& 1.46		& 9.20		&14.98		& 0.38		& 0.01		\\
sex105-1	&    7.14	& 0.76		&-1.85		& 3.46		&12.03		& 1.38		& 1.00		& 0.39		\\
sex105-2	&    2.81	& 2.70		&-2.04		& 1.83		& 8.96		& 9.33		& 0.32		& 0.01		\\
sex106-1	&    6.79	& 0.64		&-1.87		& 3.67		&11.79		& 0.97		& 1.00		& 0.50		\\
sex106-2	&    3.14	& 2.54		&-2.10		& 1.84		& 8.79		& 9.05		& 0.41		& 0.02		\\
sex107-1	&    7.53	& 0.57		&-1.83		& 4.76		&12.03		& 0.88		& 1.00		& 0.67		\\
sex107-2	&    4.67	& 1.56		&-2.05		& 2.44		& 8.57		& 3.93		& 0.89		& 0.09		\\
sex153-1	&    7.63	& 0.79		&-1.87		& 3.13		&11.90		& 1.68		& 0.99		& 0.33		\\
sex153-2	&    3.67	& 2.70		&-2.12		& 1.73		& 8.12		&18.35		& 0.46		& 0.00		\\
sex154-1	&    8.18	& 0.65		&-1.82		& 4.09		&12.20		& 1.06		& 1.00		& 0.49		\\
sex154-2	&     3.4	& 2.63		&-2.09		& 2.27		& 7.87		& 6.90		& 0.46		& 0.02		\\
sex155-1	&    7.44	& 0.61		&-1.81		& 4.10		&11.91		& 0.96		& 1.00		& 0.53		\\
sex155-2	&    4.28	& 1.59		&-2.06		& 2.50		& 7.90		& 4.52		& 0.83		& 0.09		\\
sex156-1	&    8.26	& 0.55		&-1.82		& 4.46		&12.21		& 0.91		& 1.00		& 0.62		\\
sex156-2	&    4.43	& 0.99		&-2.06		& 2.86		& 7.52		& 3.28		& 0.99		& 0.22		\\
sex157-1	&    8.99	& 0.53		&-1.83		& 4.98		&12.42		& 0.83		& 1.00		& 0.71		\\
sex157-2	&    4.49	& 0.67		&-2.05		& 4.46		& 7.68		& 0.90		& 1.00		& 0.58		\\
sex203-1	&     7.9	& 0.84		&-1.83		& 4.07		&12.36		& 1.21		& 1.00		& 0.35		\\
sex203-2	&    3.41	& 2.54		&-2.14		& 1.42		& 7.47		&64.35		& 0.51		& 0.02		\\
sex204-1	&    8.09	& 0.63		&-1.82		& 4.00		&12.41		& 1.01		& 1.00		& 0.51		\\
sex204-2	&    3.71	& 1.95		&-2.10		& 2.03		& 7.30		& 5.45		& 0.69		& 0.06		\\
sex205-1	&    9.04	& 0.58		&-1.84		& 4.58		&12.72		& 0.88		& 1.00		& 0.64		\\
sex205-2	&    3.96	& 1.23		&-2.10		& 3.21		& 6.17		& 3.96		& 0.95		& 0.17		\\
sex206-1	&    8.54	& 0.57		&-1.83		& 4.51		&12.50		& 0.85		& 1.00		& 0.64		\\
sex206-2	&    3.98	& 0.76		&-2.13		& 5.82		& 7.06		& 1.18		& 1.00		& 0.46		\\
sex207-1	&    9.46	& 0.53		&-1.84		& 4.79		&12.70		& 0.83		& 1.00		& 0.72		\\
sex207-2	&    4.87	& 0.61		&-1.98		& 4.29		& 7.17		& 0.98		& 1.00		& 0.61		\\
sex254-1	&    8.48	& 0.62		&-1.81		& 4.42		&12.61		& 0.97		& 1.00		& 0.53		\\
sex254-2	&    3.21	& 1.84		&-2.20		& 2.30		& 5.99		& 5.33		& 0.75		& 0.08		\\
sex255-1	&     9.2	& 0.56		&-1.82		& 4.63		&12.67		& 0.80		& 1.00		& 0.66		\\
sex255-2	&    3.36	& 1.09		&-2.18		& 2.85		& 5.61		& 3.43		& 0.99		& 0.21		\\
sex256-1	&    8.49	& 0.55		&-1.82		& 6.44		&12.57		& 0.84		& 1.00		& 0.69		\\
sex256-2	&    3.99	& 0.71		&-2.07		& 4.10		& 5.91		& 1.11		& 1.00		& 0.51		\\
sex257-1	&    9.61	& 0.53		&-1.82		& 4.77		&12.85		& 0.80		& 1.00		& 0.74		\\
sex257-2	&    4.38	& 0.62		&-1.99		& 5.01		&10.28		& 0.96		& 1.00		& 0.62		\\
\hline\end{tabular}
\label{tab:sexres}
\tablefoot{Properties at the present day for a Sextans type dSph in the present day. The dash after the simulation name indicates how many perigalacticon passages the dwarf has experienced, with iso indicating the isolation model. Columns are the same as in Table \ref{tab:carres}.}
\end{table*}

\begin{table*}

\caption{Sculptor~type}
\centering\begin{tabular}{lcccccccc}
\hline\hline{Model}&{$L_V$~[$10^{5}$~$L_\sun$]}&{$r_{1/2}$~[kpc]}&{$\langle$[Fe/H]$\rangle$$_{r_{1/2},{\rm RGB}}$}&{$\sigma_{v,{r_{1/2}}}$~[km~s$^{-1}$]}&{SFR$_{99}$~Gyr}&{$r_{S}$~[kpc]}&{$M_{\star}$(3~kpc)/$M_\star$}&{$M_{\rm g}(3$~kpc$)/M_{\rm g}$}\\
\hline
scl-iso	&    51.2	& 0.75 &-1.70 & 8.16 &13.94 & 1.01 & 1.00 & 0.87 \\
scl103-1	&    19.2	& 1.36		&-1.75		& 4.45		&12.35		& 7.32		& 0.87		& 0.22		\\
scl103-2	&    5.55	& 3.39		&-1.86		& 1.82		& 9.67		&39.80		& 0.15		& 0.00		\\
scl104-1	&    18.7	& 1.20		&-1.73		& 4.34		&12.10		& 8.80		& 0.94		& 0.25		\\
scl104-2	&    4.49	& 3.19		&-1.86		& 2.55		& 9.20		&31.95		& 0.14		& 0.01		\\
scl105-1	&    19.4	& 0.92		&-1.68		& 5.47		&12.10		& 5.93		& 0.99		& 0.45		\\
scl105-2	&    4.79	& 2.99		&-2.01		& 2.27		& 8.96		&19.29		& 0.18		& 0.01		\\
scl106-1	&    18.3	& 0.84		&-1.68		& 5.58		&11.87		& 1.45		& 0.99		& 0.59		\\
scl106-2	&    6.57	& 2.69		&-1.98		& 2.69		& 8.49		&15.90		& 0.29		& 0.02		\\
scl107-1	&    20.7	& 0.70		&-1.65		& 6.56		&11.88		& 1.21		& 1.00		& 0.71		\\
scl107-2	&    12.7	& 1.50		&-1.89		& 4.03		& 8.49		&10.09		& 0.87		& 0.13		\\
scl153-1	&    20.5	& 1.09		&-1.70		& 4.60		&11.91		& 3.93		& 0.93		& 0.31		\\
scl153-2	&    5.68	& 3.08		&-2.01		& 2.19		& 8.02		&67.79		& 0.21		& 0.01		\\
scl154-1	&    19.2	& 0.92		&-1.66		& 4.94		&11.64		& 6.73		& 0.99		& 0.39		\\
scl154-2	&    8.19	& 2.71		&-2.00		& 2.61		& 7.82		&44.04		& 0.36		& 0.03		\\
scl155-1	&    21.1	& 0.83		&-1.67		& 5.40		&11.66		& 2.23		& 1.00		& 0.58		\\
scl155-2	&      11	& 1.96		&-1.94		& 3.52		& 7.59		&19.26		& 0.65		& 0.06		\\
scl156-1	&    25.7	& 0.73		&-1.65		& 5.92		&12.89		& 1.20		& 1.00		& 0.69		\\
scl156-2	&    12.5	& 1.43		&-1.92		& 4.11		& 7.10		& 9.39		& 0.89		& 0.13		\\
scl157-1	&    31.7	& 0.77		&-1.67		& 6.92		&13.39		& 1.14		& 1.00		& 0.76		\\
scl157-2	&    12.8	& 1.04		&-1.91		& 5.42		& 6.84		& 4.33		& 0.99		& 0.47		\\
scl203-1	&      21	& 1.00		&-1.68		& 5.71		&12.56		& 4.50		& 0.97		& 0.33		\\
scl203-2	&    6.37	& 2.84		&-2.00		& 2.38		& 6.13		&131.62		& 0.27		& 0.02		\\
scl204-1	&    22.1	& 0.87		&-1.67		& 5.48		&12.63		& 1.82		& 0.99		& 0.51		\\
scl204-2	&    9.74	& 2.27		&-1.98		& 3.30		& 5.90		&47.53		& 0.53		& 0.05		\\
scl205-1	&    21.7	& 0.82		&-1.68		& 8.97		&12.63		& 3.09		& 1.00		& 0.61		\\
scl205-2	&    12.1	& 1.42		&-1.93		& 4.14		& 5.43		& 9.04		& 0.90		& 0.15		\\
scl206-1	&    25.6	& 0.74		&-1.67		& 6.88		&12.95		& 1.46		& 1.00		& 0.71		\\
scl206-2	&    12.5	& 1.18		&-1.93		& 5.69		& 5.19		& 4.03		& 0.99		& 0.39		\\
scl207-1	&    32.5	& 0.73		&-1.68		& 6.41		&13.40		& 1.14		& 1.00		& 0.79		\\
scl207-2	&    13.8	& 0.86		&-1.92		& 7.16		& 4.99		& 1.21		& 1.00		& 0.71		\\
scl253-1	&    25.7	& 0.86		&-1.67		& 5.36		&13.18		& 2.40		& 0.97		& 0.47		\\
scl253-2	&    6.65	& 2.92		&-2.00		& 2.54		& 4.01		&154.41		& 0.28		& 0.02		\\
scl254-1	&    22.3	& 0.88		&-1.68		& 6.35		&12.72		& 3.51		& 0.99		& 0.55		\\
scl254-2	&    10.2	& 1.97		&-2.03		& 4.14		& 3.77		& 8.07		& 0.64		& 0.07		\\
scl255-1	&    26.2	& 0.74		&-1.66		& 5.85		&13.00		& 1.23		& 1.00		& 0.69		\\
scl255-2	&      12	& 1.48		&-1.94		& 4.19		& 3.30		& 6.05		& 0.91		& 0.17		\\
scl256-1	&    26.1	& 0.72		&-1.67		& 6.33		&12.92		& 1.15		& 1.00		& 0.75		\\
scl256-2	&    12.4	& 1.10		&-1.96		& 5.77		& 3.15		& 2.49		& 0.99		& 0.44		\\
scl257-1	&    36.2	& 0.67		&-1.65		& 6.68		&13.55		& 1.07		& 1.00		& 0.82		\\
scl257-2	&    20.2	& 0.75		&-1.83		& 6.85		& 2.97		& 1.06		& 1.00		& 0.76		\\
\hline\end{tabular}
\label{tab:sclres}

\tablefoot{Properties at the present day for a Sculptor like dSph in the present day. The dash after the simulation name indicates how many perigalacticon passages the dwarf has experienced, with iso indicating the isolation model. Columns are the same as in Table \ref{tab:carres}.}
\end{table*}

\begin{table*}

\caption{Fornax~type}
\centering\begin{tabular}{lcccccccc}\hline\hline{Model}&{$L_V$~[$10^{5}$~$L_\sun$]}&{$r_{1/2}$~[kpc]}&{$\langle$[Fe/H]$\rangle$$_{r_{1/2},{\rm RGB}}$}&{$\sigma_{v,{r_{1/2}}}$~[km~s$^{-1}$]}&{SFR$_{99}$~Gyr}&{$r_{S}$~[kpc]}&{$M_{\star}$(3~kpc)/$M_\star$}&{$M_{\rm g}(3$~kpc$)/M_{\rm g}$}\\
\hline
fnx-iso	&     138	& 0.85 &-1.31 &10.75 &13.70 & 1.29 & 0.99 & 0.88 \\
fnx171-1	&     124	& 0.84		&-1.31		& 9.94		&13.59		& 1.26		& 0.99		& 0.90		\\
fnx171-2	&     118	& 0.81		&-1.33		&10.06		&13.63		& 1.34		& 0.99		& 0.87		\\
fnx201-1	&     131	& 0.83		&-1.31		&10.85		&13.68		& 1.24		& 0.99		& 0.90		\\
fnx201-2	&     121	& 0.84		&-1.37		&10.91		&13.73		& 1.17		& 1.00		& 0.89		\\
fnx251-1	&     130	& 0.84		&-1.32		&10.72		&13.68		& 1.22		& 0.99		& 0.90		\\
fnx251-2	&     132	& 0.84		&-1.36		&12.17		&13.77		& 1.12		& 1.00		& 0.91		\\
\hline\end{tabular}
\label{tab:fnxres}
\tablefoot{Properties at the present day for a Sculptor like dSph in the present day. The dash after the simulation name indicates how many perigalacticon passages the dwarf has experienced, with iso indicating the isolation model. Columns are the same as in Table \ref{tab:carres}.}
\end{table*}

}
\end{document}